\documentclass[aps,prd,11pt,superscriptaddress,floatfix,nofootinbib,notitlepage]{revtex4-1} 
\usepackage{colortbl}
\usepackage[T1]{fontenc} 
\usepackage{lmodern}
\usepackage{hyperref}
\usepackage{graphicx}
\usepackage{bm}
\usepackage{amsmath}
\usepackage{amssymb}
\usepackage{dcolumn}
\usepackage{multirow}
\usepackage{color}
\usepackage[dvipsnames]{xcolor}
\usepackage{aas_macros}
\usepackage{fancyhdr}
\usepackage[papersize={8.5in,11in},top=2cm, bottom=1.5cm, left=1.5cm, right=1.5cm]{geometry}

\begin{document}
\title{The Neutrino Signal From Pair Instability Supernovae}

\author{Warren P. \surname{Wright}}
\email{wpwright@ncsu.edu}
\affiliation{Department of Physics,  North Carolina State University, Raleigh, North Carolina 27695, USA}

\author{Matthew S. \surname{Gilmer}}
\email{msgilmer@ncsu.edu}
\affiliation{Department of Physics,  North Carolina State University, Raleigh, North Carolina 27695, USA}

\author{Carla \surname{Fr\"ohlich}}
\email{cfrohli@ncsu.edu}
\affiliation{Department of Physics,  North Carolina State University, Raleigh, North Carolina 27695, USA}

\author{James P. \surname{Kneller}}
\email{jpknelle@ncsu.edu}
\affiliation{Department of Physics,  North Carolina State University, Raleigh, North Carolina 27695, USA}

\date{\today}

\begin{abstract}
A very massive star with a carbon-oxygen core in the range of $64$~M$_{\odot}<M_{\mathrm{CO}}<133$~M$_{\odot}$ is expected to undergo a very different kind of explosion known as a pair instability supernova. Pair instability supernovae are candidates for superluminous supernovae due to the prodigious amounts of radioactive elements they create. While the basic mechanism for the explosion is understood, how a star reaches a state is not, thus observations of a nearby pair-instability supernova would allow us to test current models of stellar evolution at the extreme of stellar masses. Much will be sought within the electromagnetic radiation we detect from such a supernova but we should not forget that the neutrinos from a pair-instability supernova contain unique signatures of the event that unambiguously identify this type of explosion. We calculate the expected neutrino flux at Earth from two, one-dimensional pair-instability supernova simulations which bracket the mass range of stars which explode by this mechanism taking into account the full time and energy dependence of the neutrino emission and the flavor evolution through the outer layers of the star. We calculate the neutrino signals in five different detectors chosen to represent present or near future designs. We find the more massive progenitors explode as pair-instability supernova which can easily be detected in multiple different neutrino detectors at the `standard' supernova distance of $10\;{\rm kpc}$ producing several events in DUNE, JUNE and SuperKamiokande, while the lightest progenitors only produce a handful of events (if any) in the same detectors. The proposed HyperKamiokande detector would detect neutrinos from a large pair-instability supernova as far as $\sim 50\;{\rm kpc}$ allowing it to reach the Megallanic Clouds and the several very high mass stars known to exist there. 

\end{abstract}

\maketitle

\section{Introduction \label{Introduction}}

%
Pair instability supernovae (PISNe) are the explosive end point of very massive stars. After central carbon burning, the cores of these stars reach conditions that result in a dynamical instability due to the formation of electron-positron pairs \citep{1967PhRvL..18..379B,Ravaky.Shaviv:1967,Fraley1968}. The conversion of photons to electron-positron pairs softens the equation of state and leads to a contraction which increases the density and temperature triggering explosive burning of the oxygen. The energy release from the burning is sufficient to reverse the contraction and unbind the star in the form of a PISN explosion \citep{2002ApJ...567..532H,2012ApJ...748...42C}. The existence of such supernovae would be important from both a chemical evolution \cite{2008ApJ...679....6K,2012ApJ...745...50W} and galaxy formation \citep{2008ApJ...682...49W} perspective. In addition to the large amounts of $^{56}$Ni they produce, the predicted nucleosynthesis pattern from PISNe has a strong odd-even staggering \citep{2002ApJ...567..532H}. Such a pattern has not yet been found conclusively in any metal-poor stars \cite{Ji2015} but one tentative case has been identified \cite{Aoki2014}. 

For a star to experience the pair instability (PI), it needs to be massive enough to form a carbon-oxygen core in the range of $64$~M$_{\odot}<M_{\mathrm{CO}}<133$~M$_{\odot}$ \citep{2002ApJ...567..532H}. The corresponding zero age main sequence (ZAMS) mass is highly dependent on the details of stellar evolution and in particular on the mass loss history, which depends on the metallicity and rotation, but is not well understood \citep{Vink2015}. 
The long-standing expectation was that only non-rotating stars of zero metallicity (which are thought to have very little mass loss) born with ZAMS masses in the range of 140~M$_{\odot}< M_{\mathrm{ZAMS}}<260$~M$_{\odot}$ explode this way \citep{2002ApJ...567..532H}. Stars with higher ZAMS mass will collapse directly to a black hole; stars with an initial mass less than 140~M$_{\odot}$ may undergo a series of pair instability pulses (pulsational PISN) but ultimately explode as core-collapse supernovae. 
However it has been found that including rotation in stellar models can facilitate a chemically more homogeneous evolution. This leads to larger carbon-oxygen cores for the same initial mass, and hence shifts the mass range of those stars which undergo PISN to a lower ZAMS mass limit. In simulations at zero metallicity and with initial rotations rates of 80\% of Keplerian velocity, stars with an initial masses as low as 40~M$_{\odot}$ are found to undergo a pulsational PISN and stars above 65~M$_{\odot}$ undergo a full PISN \cite{2012ApJ...748...42C}.
Also more recently, the viewpoint that the star had to be metal free to explode as a PISN has been challenged. Langer \emph{et al.} \citep{2007A&A...475L..19L} find that PISN may occur at metallicities as high as Z$_{\odot}/3$ and simulations at solar metallicity find that a magnetic field at the surface of the massive star can reduce the mass-loss rate so that even stars at near-solar metallicity can explode as PISN \cite{Georgy.Meynet.ea:2017}. Thus it appears there is still a great deal we do not understand about exactly which stars explode as PISN. Observations of PISN would be of great value in answering the numerous questions about the progenitors and, if PISN can occur at non-zero metallicites, then it may be possible to observe a nearby event from which we could hope to obtain high quality information. 

The theoretical expectation for the rate of PISN is very uncertain because of the aforementioned uncertain mass loss of massive stars. The estimate by Langer \emph{et al.} \cite{2007A&A...475L..19L} is that there is one PISN event for every 1000 SNe in the local universe. 
To make this estimate the authors considered only the mass range 140~M$_{\odot}< M_{\mathrm{ZAMS}}<260$~M$_{\odot}$ and thus do not include the previously mentioned extension down to lower ZAMS masses in the presence of rapid rotation. Additionally, while they include the effects of magnetic torques on mixing, they do not consider a surface dipolar magnetic field and its affect on the mass-loss rate, as in \citet{Georgy.Meynet.ea:2017}. 
Due to this uncertainty, we adopt a lower limit for the minimum ZAMS mass of PISN progenitors of 100~M$_{\odot}$. If that is the case, one can find many candidates of potential PISN progenitors within the Milky Way and nearby galaxies. Several stars with masses exceeding 100~M$_{\odot}$ have been found in the Magellanic Clouds \citep{Kudritzki.ea:1996}. Crowthers \emph{et al.} \citep{2016MNRAS.458..624C} identified almost two dozen stars in R136/30 Dor whose initial masses exceed $\sim100$~M$_{\odot}$. Very massive stars are also known in NGC3603 \citep{2008MNRAS.389L..38S} and the Arches cluster \citep{2008A&A...478..219M}. Some of these stars exceed the previously assumed upper mass limit of $\sim 150\;{\rm M_{\odot}}$ for stars in metal-rich galaxies \citep{2005Natur.434..192F,2006MNRAS.365..590K}. If one of these stars were to explode as a PISN, one wonders how such an event might be identified.

There are several observational features which might be used to identify a PISN. In recent years, a number of authors have revisited the pair-instability mechanism computing expected lightcurves for various PISN models from zero metallicity to $10^{-3}\,$Z$_{\odot}$ \cite{2011ApJ...734..102K,Whalen.CosmicExplI:2013,Whalen.CosmicExplIII:2014,2017MNRAS.464.2854K}. These studies have found the peak luminosity in the models can be comparable to the peak luminosity of observed superluminous supernovae.
Indeed a few of the observed superluminous supernovae with slower decay times have been discussed as possibly being of PISN origin: for example, the energy required to power the lightcurve of SN~2006gy (at a distance of $73\;{\rm Mpc}$) could be provided by the decay of 22~M$_{\odot}$ of $^{56}$Ni \cite{0004-637X-666-2-1116}; SN~2007bi has been suggested to be a pair-instability supernova by Gal-Yam \emph{et al.} \cite{2009Natur.462..624G}, and another candidate  was found by Cooke \emph{et al.} \cite{2012Natur.491..228C}. Superluminous supernova PS1-14bj also has a slowly evolving light curve, consistent with PISN models \cite{Lunnan.ea.PS1-14bj:2016}. Of course, in addition to the energy from the decay of the large of amount of $^{56}$Ni, other processes, such as interactions of the ejecta with a circum-steller medium, could also contribute to the high luminosity.
However we note that the calculated PISN lightcurves at non-zero metallicity are generally dimmer than the lightcurves at zero metallicity so PISN in the local Universe may become hidden among other classes of supernovae, see e.g.\ \cite{2014ApJ...797....9W} or \cite{2014A&A...566A.146K}. Should a nearby PISN occur, one would expect the increased quality of the observations might be able to find enough differences between a PISN and other supernova types to distinguish them but, as always, extracting information from the electromagnetic signal requires a high degree of modeling of the ejecta material which frequently introduces some uncertainty. 

A much cleaner way to unambiguously distinguish PISN from other types of stellar explosions is the neutrino signal because the neutrino emission mechanism for PISN is the same as for Type Ia supernovae \cite{kunugise2007a,Odrzywolek2011a,Seitenzahl2015a,2016PhRvD..94b5026W,2017PhRvD..95d3006W} but on a much larger scale. The aim of this paper is to compute the expected neutrino signal from a PISN and determine its detectability in current and near-future neutrino detectors in order to answer the question of whether the neutrinos can be added to the suite of tools one can use to identify a PISN. We take into account the full time and energy dependence of the emission and the follow the neutrino flavor transformation in the envelope with a complete 3-flavor oscillation code. 

Our paper is arranged as follows. In Sec. \ref{sec:SneModel} we briefly describe the simulation of the pair-instability supernova used to compute the neutrino signal. In Sec. \ref{sec:Production} we outline the algorithm for computing the neutrino spectrum and luminosity from the simulation and then in Sec. \ref{sec:NeutrinoOscillation} we present the calculation of the flavor transformation through the envelope. We then consider (Sec. \ref{sec:NeutrinoDetection}) a suite of neutrino detectors in order to determine how well, and to what distance, a pair-instability supernova could be detected. Section \ref{sec:AstroContext} discusses the availability of nearby high mass stars and we conclude in Sec. \ref{sec:Conclusion}.

\section{Supernova Simulation \label{sec:SneModel}}

For the PISN simulation we use two GENEC \citep{Ekstrom2012,Yusof2013} progenitor models, P250 and P150 \citep{2017MNRAS.464.2854K,Gilmer2017}. As their names suggest, the initial masses of these models are 250~M$_{\odot}$ and 150~M$_{\odot}$, respectively. Throughout the rest of our paper we shall distinguish our calculations by the progenitor model. The models have initial metallicity Z = 10$^{-3}$ and were evolved without rotation. The Schwarzschild criterion was used for convection and the outer convective zone was treated according to the mixing length theory with $\alpha_{MLT} = 1.0$ ($\alpha_{MLT} = l/H_{\rho}$, where $l$ is the mixing length and $H_{\rho}$ is the density scale height). Nearing the end of core carbon burning (after 25,000--30,00 time steps; 300--800 mass zones), when a sufficiently large part of the core has become unstable, the models are ready to explode. At this point the carbon-oxygen core mass of P250 is 126.7 M$_{\odot}$ which is near the upper end of the PISN mass range. The P150 model has a carbon-oxygen core mass of 65.7~M$_{\odot}$ which is at the lower end of the PISN mass range. Thus the neutrino emission we calculate from these two models should bracket what we can expect from an actual PISN.

To model the explosion phase, we used version 4.3 of the FLASH multiphysics code \citep{Fryxell2000,Dubey2009}. We used a subset of standard code modules suitable for the present application including the Helmholz EOS and the Aprox19 nuclear reaction network \citep{Timmes.Swesty:2000}, the directionally unsplit hydrodynamics solver \citep{LeeDeane2009}, the improved multipole solver for self-gravity \citep{Couch2013}, and its adaptive meshing capability. We run on a 1D grid of spherical geometry with an effective mesh resolution of 1.3 $\times$ 10$^{8}$ cm. The grid has the freedom to refine once (a local doubling of resolution) according to the density and temperature gradients. We set the refinement cutoff values to half of their default values. We consider only the innermost 5 $\times$ 10$^{10}$ cm of the GENEC progenitor models because the outer material will not contribute to either the explosion dynamics, neutrino emission or neutrino flavor oscillation. The boundary conditions used are `reflect' and `diode' (a modification of `outflow' in which matter is prevented from flowing into the grid) for the inner and outer boundaries, respectively. We then evolve the models through the contraction and explosion phases until nucleosynthesis is complete. Additional details on the progenitor models and the explosion calculation can be found in \citet{Gilmer2017}. The thermodynamic trajectory of the core then serves as an input for the neutrino emission calculation.

\section{Neutrino Production \label{sec:Production}}
The strategy used to calculate the neutrino emission from our PISN simulations is similar to the one used in \cite{2016PhRvD..94b5026W} and \cite{2017PhRvD..95d3006W} for the case of Type Ia SN. We use the software package \textsc{NuLib} \cite{Fuller1982,Langanke2000,Langanke2003,Burrows2006a,Oda1994,Steiner2013,OConnor2010,Sullivan2015} to calculate the emission as a function of time, neutrino energy and flavor and due to the different emission processes. The weak processes included are electron and positron capture on neutrons, protons and nuclei. The only thermal process included is neutrino pair creation due to electrons annihilating with positrons. This emission calculation is done by post-processing the simulations described in the previous section. Density, temperature, electron fraction and isotopic composition are extracted from each simulation for each time-slice and radial zone. These quantities are used to setup an Equation of State (EOS) which is then used by \textsc{NuLib} to compute neutrino emissivity. For comparison, two different EOSs have been considered. The first is the Helmholtz equation of state (based on \cite{Itoh1996a}), the second is the SFHo EOS \cite{Steiner2013}. The strength of the Helmholtz EOS is consistency, because it is the same EOS used in the original FLASH simulation. However, the simulation (with the Helmholtz EOS and the Aprox19 nuclear reaction network \citep{Timmes.Swesty:2000}) does not track all isotopes and thus neutrino emission from weak processes on isotopes not included, is missing. Nonetheless, given that the dominant isotopes are tracked, it is expected that these missing nuclei would not account for the bulk of neutrino emission from weak processes but it might miss spectral features especially towards higher neutrino emission energies. To test this hypothesis we consider a second EOS, SFHo, which assumes Nuclear Statistical Equilibrium (NSE) thus allowing us to a include emission from a wider range of nuclei. This EOS is an accurate description of the composition of material in a core-collapse supernova (CCSN) when $T>0.5\text{ MeV}\approx5.8\text{ GK}$. However, our PISN simulation does not get as hot or dense as a CCSN, and thus our strategy, when using the SFHo EOS, is to only compute the weak emission from computational zones with $T>3\text{ GK}$ but when we calculate the thermal emission with this EOS we include zones with $T>1.2\text{ GK}$, which is approximately the lower bound of the SFHo EOS (see \cite{Steiner2013}). Our choice for the cutoff of $T>3\text{ GK}$ for zones from which we calculate the emission due to weak processes was so that we only include zones hot and dense enough to be close to NSE, while avoiding using a NSE EOS in zones that should not be in equilibrium. This strategy was proven to be successful when applied to the case of Type Ia SN \cite{2016PhRvD..94b5026W} and \cite{2017PhRvD..95d3006W}.

\begin{figure}[ht]
\includegraphics[width=0.75\linewidth]{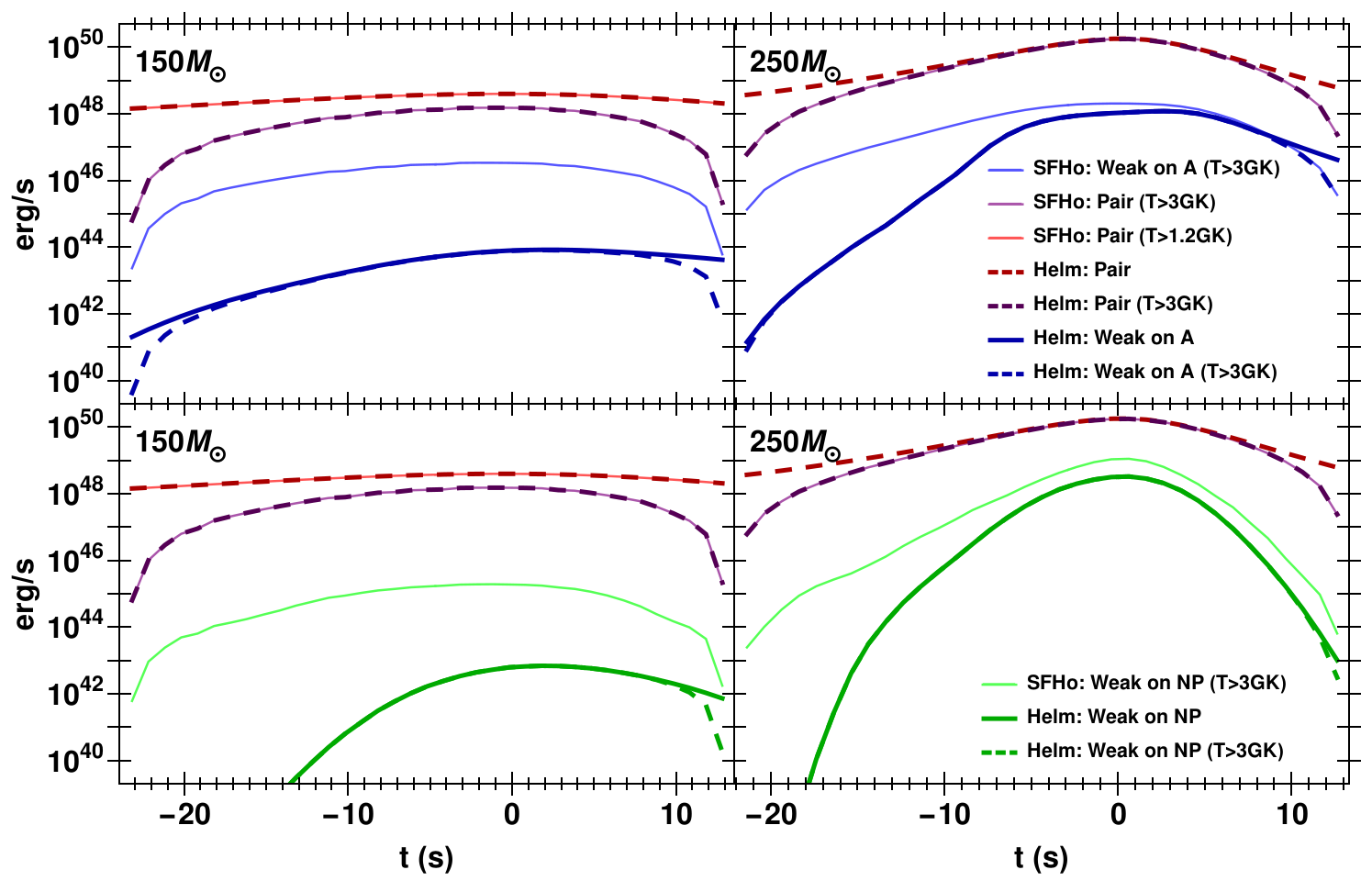}
\caption{PISN total neutrino luminosity as a function of time arising from the various neutrino emission processes considered. The results for the P150 simulation are shown on the left, the P250 on the right. Emission due to pair production are shown as red and purple lines, electron capture on nuclei are the blue lines, and and electron and positron capture on nucleons are the green lines. The various temperature cuts used are given in the legend. We have also calculated the weak and thermal emission using the Helmholtz EOS using temperature cutoffs in order to make a comparison although this EOS does not fail at lower temperatures in the same way as the SFHo EOS.}  
\label{fig:TotalNeutrinoLuminosityVStime}
\end{figure}

Figure \ref{fig:TotalNeutrinoLuminosityVStime} shows the results of the neutrino emission calculations using \textsc{NuLib}. The time denoted as $t=0$ s is the time of maximum compression which is determined by locating the minimum in the gravitational energy. It is no surprise that the time of maximum compression is also the time of maximum neutrino emission. This is because increasing compression generally causes an increase in temperature which itself causes greater thermal activity, including neutrino emission. It is worth emphasizing that the neutrino signal from a PISN has a long duration, $\sim30$ s, much longer than that of Type Ia supernovae \cite{Odrzywolek2011a,2016PhRvD..94b5026W,2017PhRvD..95d3006W} and even core-collapse supernovae \cite{1998ApJ...496..216T,2010A&A...517A..80F}. 
In this figure, pair production (all six neutrino flavors), electron capture on nuclei ($\nu_e$), and electron and positron capture on nucleons ($\nu_e$ and $\bar{\nu}_e$) are given in red, blue and green respectively (positron capture on nuclei is very subordinate). The results for the P150 (left) and the P250 (right) models are presented as well as the results for the Helmholtz EOS and the SFHo EOS. Additionally, results when the Helmholtz EOS is assumed and a $T>3$ GK cutoff is imposed, are also displayed. For P250, there is no thin red line representing SFHo pair emission with $T>1.2$ GK because for this SN, a $T>3$ GK cut-off still captures the vast majority of the signal (unlike in the P150 case). In all cases we see that thermal (pair) emission dominates. 
For P150, pair emission using SFHo with a $T>1.2$ GK cut-off agree with the Helmholtz results with no cut-off. Also in agreement are the SFHo and Helmholtz pair emission results when both use a $T>3$ GK cut-off. For P250 at times during the peak of the neutrino emission, pair emission using SFHo with a $T>3$ GK cut-off agree with the Helmholtz results with no cut-off and those with a $T>3$ GK cut-off. Thus we conclude that EOS choice has little effect on neutrino pair emission, which is the dominant emission process. We can also conclude that the use of a temperature cut-off needs to be done carefully: using a cut-off temperature which is too high can lead to substantial error in the emission calculations especially for the for the cooler (lower mass) PISN.

For both electron capture on nuclei and electron and positron capture on nucleons, we again see that the temperature cut-off has a much small effect. This is clear because the thick blue and green lines match very well with their respective dashed lines, especially around times of peak emission for both the P150 simulation and the P250 simulation. However, the choice of EOS does make a large difference for this emission processes. We observe that the thin blue and green lines are above the thick or dashed blue and green lines, especially at early times. The difference indicates the SFHo EOS, which assumes NSE, over-predicts the amount of weak neutrino emission. Given that the P150 simulation barely reaches temperatures above the $T>3$ cut-off, the large over prediction for an EOS which assumes equilibrium has been reached is not surprising. However for both simulations, it is clear that the differences between the EOSs decrease as the explosions progress. If the mismatch at early times can be attributed to the SFHo predicting neutrino emission from nuclei that have not yet been synthesized then, later on when those nuclei have indeed been formed, the assumption of NSE is closer to reality and the SFHo EOS is more accurate. 
While these differences in the weak process emission are interesting, we stress that thermal emission is the dominant emission process in both simulations and because this emission can be calculated accurately, the mismatches in the weak emission only contribute to the uncertainty in sub-leading features of the spectrum. 

\begin{figure}[ht]
\includegraphics[width=0.7\linewidth]{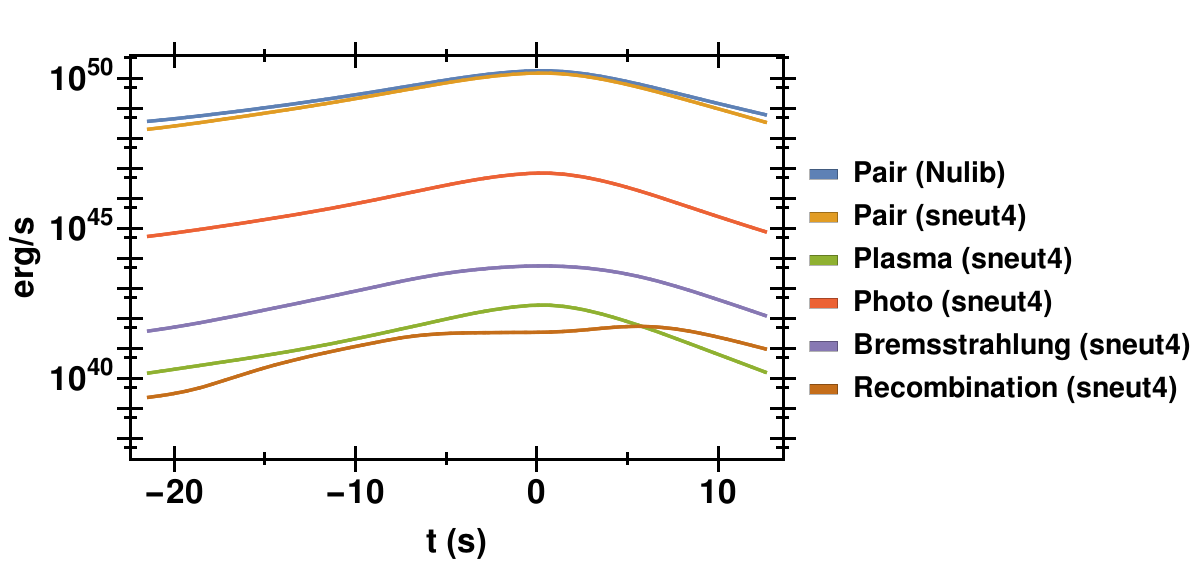}
\caption{PISN total neutrino luminosity for thermal processes for the P250 simulation. The different thermal processes are color coded as per the legend and the code used to generate each is given in parentheses. The calculation assumes the Helmholtz EOS for all processes. The "Pair (Nulib)" line is a reproduction of the "Helm: Pair" line from Fig. \ref{fig:TotalNeutrinoLuminosityVStime}.}
\label{fig:TotalNeutrinoLuminosityVStimeItoh}
\end{figure}

That pair production is the dominant thermal process for a PISN is shown in Fig. \ref{fig:TotalNeutrinoLuminosityVStimeItoh} which shows the all-flavor neutrino luminosity for a number of thermal processes computed for the P250 simulation. The thermal processes included in the calculation are pair, photo-, plasma, bremsstrahlung and recombination neutrino processes and were calculated using the code \textsc{sneut4} which is based on \cite{Itoh1996a} which can be found at \url{http://cococubed.asu.edu/code_pages/nuloss.shtml}. These calculations have no spectral information but allow us to determine which processes are important. Figure \ref{fig:TotalNeutrinoLuminosityVStimeItoh} shows that pair production is by far the most important thermal process at all times. Also, the pair results from \textsc{NuLib} very nearly overlap with the pair results from \textsc{sneut4}, giving further confidence to our results. A similar analysis was done for the P150 simulation and the conclusion that pair production is by far the most important thermal process remains true.

\begin{figure}[ht]
\includegraphics[width=0.8\linewidth]{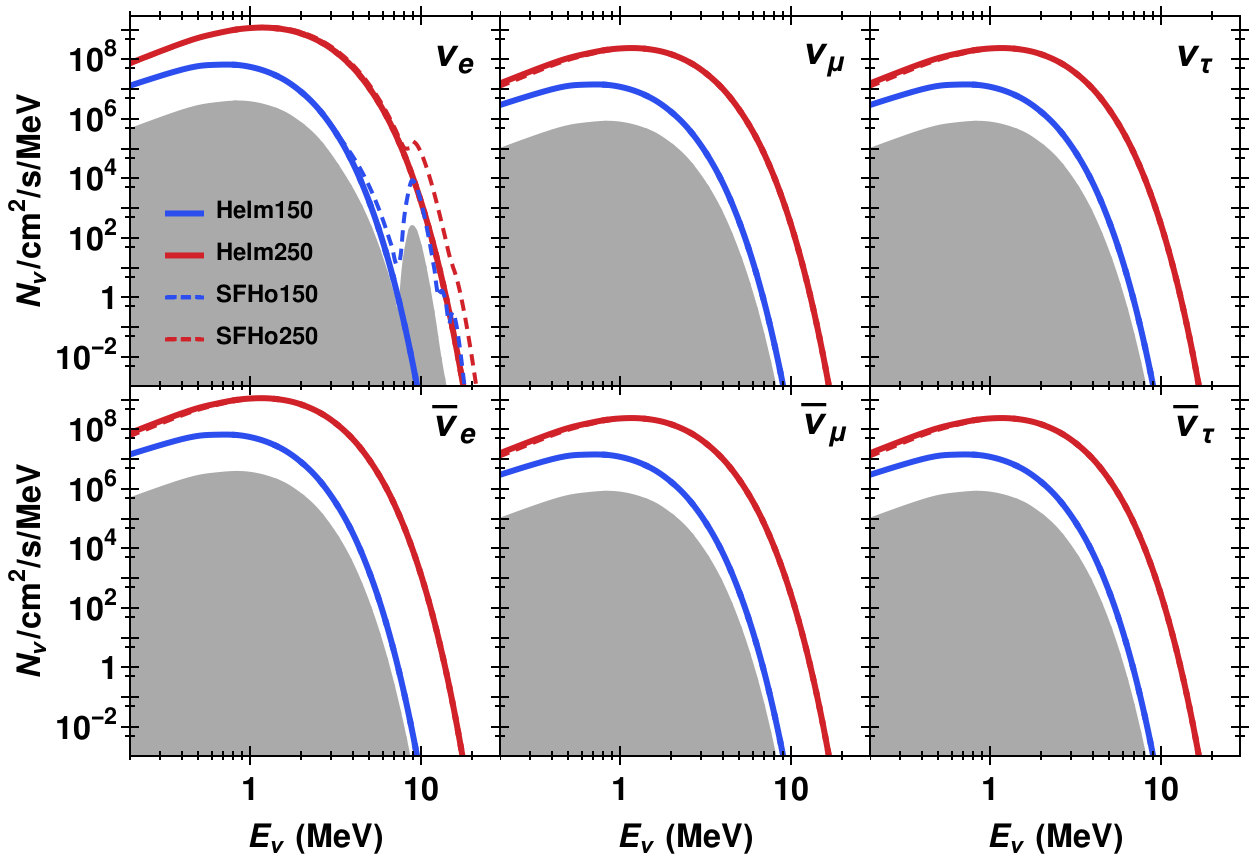}
\caption{PISN neutrino flux spectra. Each curve is the sum of all considered weak and thermal processes at the timeslice of maximum emission. The gray region is the spectra from the SFHo, P250 results at $t=12.6$ s.}
\label{fig:UnOscFlux}
\end{figure}

The advantage the \textsc{NuLib} calculations have is that they contain spectral information. Figure \ref{fig:UnOscFlux} shows the neutrino spectrum as a function of neutrino energy for all six flavors and for the time slice around maximum emission. The red (blue) lines represent the P250 (P150) simulation and the solid and dashed lines represent the results using the Helmholtz and SFHo EOSs respectively. 
The figure shows, that for all flavors other than $\nu_e$, the spectra have the same expected thermal shape with a peak at $\sim1.5$ MeV for the P250 simulation and $\sim1$ MeV for the P150 simulation. The $\nu_e$ spectra has a strong contribution from weak emission which boosts the overall rate and additionally introduces a spectral feature at 10 MeV when we use the SFHo EOS.  
Given the results from \cite{2016PhRvD..94b5026W} and \cite{2017PhRvD..95d3006W}, we recognize this 10 MeV peak as weak emission mostly from electron capture on copper. We find that this 10 MeV peak is present at all time slices (for the SFHo results). However, given the aforementioned mismatch between the weak process emission using the two EOSs, our confidence in the presence of this feature in the spectrum is low initially and increases as the simulation progresses (especially for the P250 simulation). 

Finally, the time evolution of these spectra are simply described as brightening then fading with little change of the shape except for the spectral feature around 10 MeV in $\nu_e$ which becomes relatively more pronounced later in the simulation. This behavior is demonstrated by the gray region which is the spectra from the SFHo, P250 results at $t=12.6$ s.
Another interesting point is that the thermal peak energy does not change much as the supernova progresses. For the Type Ia supernova analysis previously performed with this same strategy, the thermal peak shifted to lower energies as the explosion progressed implying that each zone in the Type Ia supernova cooled over time. The different evolution for the PISN indicates that the hot zones remain hot for a much longer period of time than in the Type Ia simulations. 


\section{Neutrino Flavor Transformation\label{sec:NeutrinoOscillation}}
In order to accurately determine the neutrino signal from a PISN that reaches Earth, the effects of neutrino oscillation need to be accounted for. This includes both the neutrino flavor oscillations that take place during neutrino propagation through the stellar mantle and the decoherence that arises as the neutrino propagates through the vacuum between the supernova and Earth. The calculation strategy we have used for the flavor transformations is the same as the one used in \cite{2016PhRvD..94b5026W} and \cite{2017PhRvD..95d3006W}.

Neutrino oscillation phenomena are calculated by solving the Schr\"{o}dinger equation. The Schr\"{o}dinger equation in some basis $(X)$ is
\begin{equation}
\imath \frac{dS^{(XX)}}{dr} = H^{(X)} S^{(XX)} \label{eqn:Schroedinger},
\end{equation}
where $H^{(X)}$ is the Hamiltonian in that basis and the evolution matrix $S^{(YX)}(r_2,r_1)$ connects the neutrino states in the $(X)$ basis at some initial position $r_1$ to the states in a possibly different basis $(Y)$ at $r_2$. Note that when we are not referring to a specific basis, we shall drop the superscript. The evolution of the antineutrinos is governed by an evolution matrix $\bar{S}$ which evolves according to the same equation but with a Hamiltonian $\bar{H}$. From the evolution matrix we define the transition probability $P^{(YX)}_{yx} = \left|S^{(YX)}_{yx}\right|^2$ as the probability for some state $x$ in the $(X)$ basis at $r_1$ to be in state $y$ in the $(Y)$ basis at $r_2$. Antineutrino transition probabilities will be denoted with an overbar. The two bases commonly used in the literature are the flavor basis -- which has basis states $\nu_{e}$, $\nu_{\mu}$, and $\nu_{\tau}$ -- and the matter basis (often called the mass basis when in vacuum) -- which has basis states $\nu_{1}$, $\nu_{2}$, and $\nu_{3}$ \cite{1986PhRvL..56.1305B}. For future reference, we shall use Greek letters $\alpha$ and $\beta$ to denote generic flavor states and the Roman symbols $i$ and $j$ to denote generic matter basis states. Again, we denote the antineutrino states in the two bases by an overbar. Using this notation, the neutrino transition probability for a neutrino that was initially state $j$ in the matter basis to be detected as state $i$ in the matter basis is denoted by $P^{\text{(mm)}}_{ij} = P(\nu_j \rightarrow \nu_i)$. 

The mixing matrix $U_V$ which describes the transformation of the flavor basis to the mass basis is parameterized by three angles $\theta_{12}$, $\theta_{23}$ \& $\theta_{13}$, and a CP-violating phase $\delta_{CP}$, and can be written as
\begin{align*}
		U_V=
		\begin{pmatrix}
		1 & 0 & 0 \\
		0 & c_{23} & s_{23} \\
		0 & -s_{23} & c_{23}
		\end{pmatrix} 
		\begin{pmatrix}
		c_{13} & 0 & s_{13}e^{-i\delta_{CP}} \\
		0 & 1 & 0 \\
		-s_{13}e^{i\delta_{CP}} & 0 & c_{13} 
		\end{pmatrix}
		\begin{pmatrix}
		c_{12} & s_{12} & 0 \\
		-s_{12} & c_{12} & 0 \\
		0 & 0 & 1
		\end{pmatrix},
	\end{align*}
with $c_{ij}=\cos\theta_{ij}$ and $s_{ij}=\sin\theta_{ij}$.

\subsection{Vacuum Hamiltonian}

In order to solve the Schr\"{o}dinger equation we need to define the Hamiltonian $H$ ($\bar{H}$ for the antineutrinos). In the vacuum, $H$ is given by a single matrix $H_V$ whose exact structure depends upon the basis one is using. In the mass basis the three neutrino states have masses $m_1$, $m_2$ and $m_3$ and the vacuum Hamiltonian is diagonal. The vacuum Hamiltonian in the flavor basis, denoted by $H_V^{(f)}$, is 
\begin{equation}
H^{(f)}_V = \frac{1}{2E}\,U_V \begin{pmatrix}
		m_1^2 & 0 & 0 \\
		0 & m_2^2 & 0 \\
		0 & 0 & m_3^2
		\end{pmatrix} U_V^{\dagger},
\end{equation}
with $E$ the neutrino energy. The mixing matrix angles and phases, together with the squared differences between neutrino masses, are part of what determines the nature of the oscillation phenomenology. In this paper we adopt the following values for all results
\begin{align}
\left( m^2_2-m^2_1,|m^2_{3}-m^2_{2}|,\theta_{12},\theta_{13},\theta_{23},\delta_{CP}\right)=\left(7.5\times10^{-5}\text{eV}^2,2.32\times10^{-3}\text{eV}^2,33.9^\circ,9^\circ,45^\circ,0\right).
\end{align}
where $m^2_{3}-m^2_{2}>0$ is for normal mass ordering (NMO) and $m^2_{3}-m^2_{2}<0$ is for inverted mass ordering (IMO). For the antineutrinos the vacuum Hamiltonian $\bar{H}_V$ is simply the complex conjugate of $H_V$. 


\subsection{Matter Hamiltonian}

In matter we must add to $H_V$ an additional term often known as the `matter Hamiltonian', $H_M$, so that $H=H_V+H_M$. In the flavor basis $H_M^{(f)}$ is given by $H^{(f)}_M = \sqrt{2}\,G_\text{F}\,n_e\,\text{Diag}\left(1,0,0\right)$, where $n_e$ is the neutrino density and $\text{Diag}\left(1,0,0\right)$ is a 3-by-3 matrix where the only non-zero entry is unity in the first diagonal element. For the antineutrinos the matter Hamiltonian $\bar{H}_M$ is related to $H_M$ by $\bar{H}_M = -H_M$.

\begin{figure}[ht]
\includegraphics[width=0.9\linewidth]{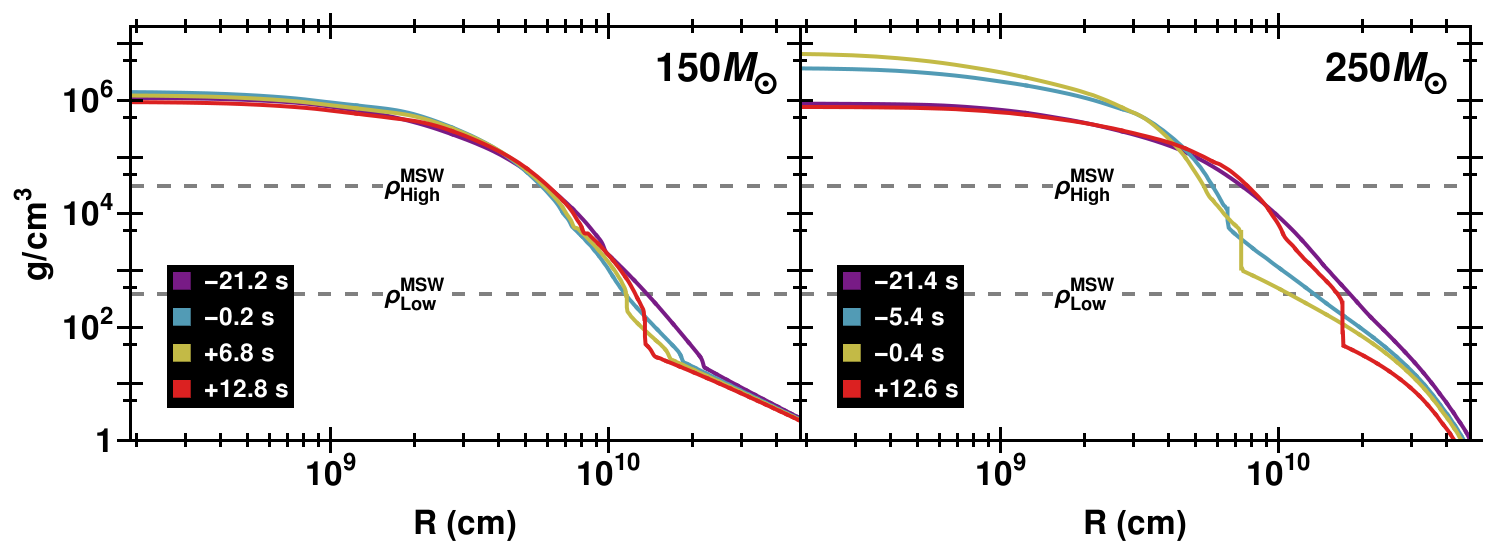}
\caption{Density profiles for various time slices for P150 (left) and for P250 (right). The two horizontal dashed lines are the 2-flavor MSW densities for $1\;{\rm MeV}$ neutrinos. The density of the high-density MSW resonance, $\rho^{MSW}_{High}$, we use $\delta m_{32}^2$ for the mass splitting and $\theta_{13}$ for the mixing angle, while for the low-density MSW resonance $\rho^{MSW}_{Low}$, we use $\delta m_{12}^2$ for the mass splitting $\delta m^2$ and $\theta_{12}$ for the mixing angle $\theta$.}
\label{fig:Density}
\end{figure}

Thus we see neutrino flavor transformation depends on the density of the material and its electron fraction $Y_e$. For the PISN simulations considered here, the electron fraction is very nearly $Y_e = 0.5$ for the entire duration of significant neutrino emission. The stellar density profile is shown in Fig. \ref{fig:Density}. The time slices chosen are the first, peak, last and the first time slice with a shock present. We see that the star develops a shock at $\sim7\times10^9$ cm at $t\approx-5.4$ s and $\sim10^4$ g/ccm for the P250 simulation and similarly at $\sim10^{10}$ cm at $t\approx6.8$ s and $\sim10^2$ g/ccm for the P150 simulation. This shock then propagates outward into regions of lower density. The densities at which MSW resonances occur for 1 MeV neutrinos are indicated by the dashed lines. To calculate the density of the high-density MSW resonance, $\rho^{MSW}_{High}$, we use the two-flavor approximation 
\begin{equation}
\rho^{MSW} = \frac{m_N}{\sqrt{2}\,G_F\,Y_e}\,\left| \frac{\delta m^2\,\cos 2\theta}{2\,E} \right| \label{eq:rhoMSW}
\end{equation}
and $\delta m_{32}^2$ for the mass splitting $\delta m^2$ in the formula and $\theta_{13}$ for the mixing angle $\theta$, while for the low-density MSW resonance $\rho^{MSW}_{Low}$, we use $\delta m_{12}^2$ for the mass splitting and $\theta_{12}$ for the mixing angle. 
In Eq. \ref{eq:rhoMSW} $m_N$ is the nucleon mass. 


\subsection{Neutrino self interaction potential\label{sec:Hnunu}}
Another possible contribution to neutrino flavor oscillation arises due to neutrino self-interaction, i.e.\ we must add a third term to $H$ (see \cite{2009JPhG...36k3201D,2010ARNPS..60..569D,Mirizzi2015a} for reviews where neutrino self interaction is discussed in core-collapse supernovae). Naively, one might not anticipate this contribution to be important in a PISN because the neutrino densities are quite low. However, the matter density - shown in Fig. \ref{fig:Density} - is also much lower than one finds in core-collapse supernovae and thus it is not immediately obvious whether or not neutrino self-interactions can be ignored. 
\begin{figure}[ht]
\includegraphics[width=0.95\linewidth]{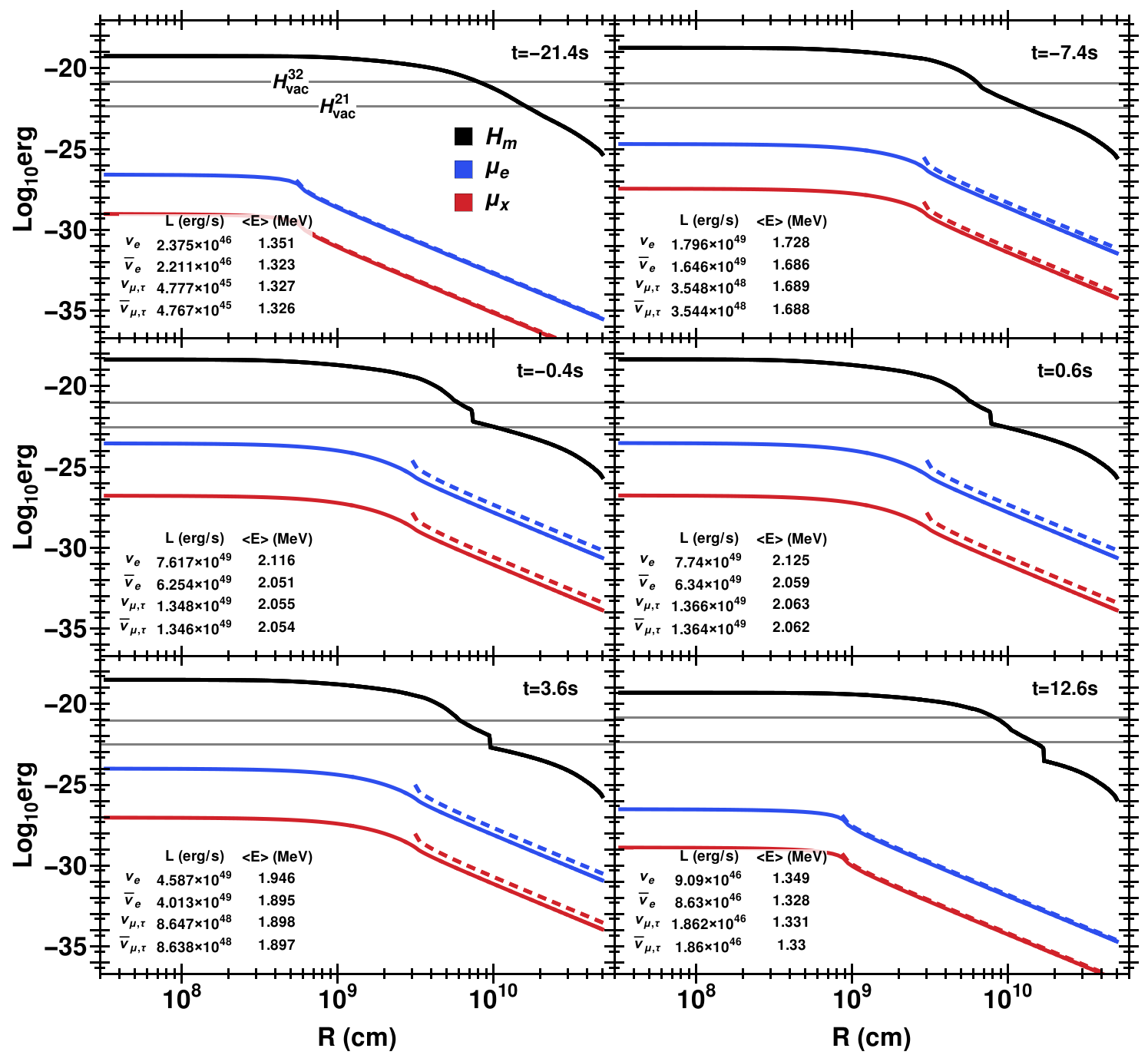}
\caption{The various Hamiltonian contributions for time slices around peak emission. Black denotes the matter Hamiltonian, gray horizontal lines are the vacuum potentials, blue are the electron component of the self-interaction potential and red the muon or tau self-interaction potential. The results are for the P250 simulation.}
\label{fig:Potentials}
\end{figure}
In order to check whether self-interaction effects might occur, we compute the 
strength of the neutrino self-interaction upon a radially emitted test neutrino. 
One important detail we need to include in our estimate is that the matter in a PISN is nowhere optically thick to neutrinos. Given that our simulation is one dimensional, our first step is to map the density and neutrino luminosity onto a sequence of spherical shells labeled by an integer $i$ and with a thickness $\delta r_i$ given by the radial grid spacing. For shell $i$, the neutrino luminosity emitted in flavor $\alpha$ per unit volume is $l_{\alpha,i}=L_{\alpha,i}/V_i$ where $V_i$ is the shell volume and $L_{\alpha,i}$ the luminosity of the shell in neutrino flavor $\alpha$. Because the matter densities are much too low to trap neutrinos, the neutrinos are emitted isotropically from each volume element of a shell. Thus, for a test neutrino emitted radially at a distance $R$ from the center of the star, the contribution to the neutrino self interaction potential (not including oscillations) from neutrinos emitted from a volume element which is at a distance $r_i$ from the center of the star and at an angle $\theta$ with respect to the ray of the test neutrino, is
\begin{align}
	d\mu_{\alpha,i,\theta} = \sqrt{2}\,G_\text{F}\frac{(1-\cos\Theta)}{4\pi c \ell^2 }
    \left(
    	\frac{l_{\alpha,i}}{\left\langle E_{\alpha}\right\rangle}
        -
        \frac{l_{\bar{\alpha},i}}{\left\langle E_{\bar{\alpha}}\right\rangle}
    \right)2\pi\sin\theta\, r_i^2 \delta r_i d\theta \label{eq:mu}
\end{align}
where $G_\text{F}$ is the Fermi constant, $\Theta$ is the angle between the ray of the test neutrino and the neutrinos emitted from the volume element, $\ell$ is the distance between the volume element and the test neutrino, $\left\langle E_{\alpha}\right\rangle$ is the average neutrino energy and $\bar{\alpha}$ represents the antineutrino of flavor $\alpha$. The distance $\ell$ is simply
\begin{equation}
\ell^2 = r_i^2 + R^2 - 2\,r_i \,R\cos\theta
\end{equation}
and the angle $\Theta$ is
\begin{align}
\cos\Theta&=\frac{R - r_i\,\cos\theta}{\ell}
\end{align}
A particular shell's contribution to the total neutrino luminosity is assumed to scale as $L_{\alpha,i}\propto T_i^9$ in accordance with the results derived in Burrows, Reddy, \& Thompson \cite{Burrows2006a} for pair production neutrinos. We determine the proportionality constant by constraining the sum over shells to give the total luminosity we have previously calculated. The contribution from an entire shell can be obtained by performing the integration over the angle $\theta$ and the contribution from the whole star can be obtained by summing over all shells. In this way we calculate the total strength $\mu_{\alpha}$ of the neutrino self-interaction.

Our temperature cut-off strategy when computing the emission means that if a shell has a temperature below the cut-ff then its neutrino luminosity is zero. Thus, for each time-slice, only shells within a certain radius emit neutrinos. This outermost neutrino-emitting shell is not a neutrino-sphere because the neutrinos are not all emitted from its surface (which is the case for a neutrino-sphere in CCSN). Nonetheless, it is instructive to treat the outermost shell as a neutrino-sphere and compute the self-interaction potential via the usual equations in the neutrino BULB model (i.e.\ via Eqn. 40c taken from \cite{2006PhRvD..74j5014D}) to compare to our shell calculation. These pure BULB model results are an upper limit for our shell model calculations because, by assuming all neutrino luminosity comes from the outermost emitting shell, there is greater flux of neutrinos from large angles $\Theta$ and thus the self-interaction is artificially increased via the factor $\left(1-\cos\Theta\right)$ in Eqn. \ref{eq:mu}.

Figure \ref{fig:Potentials} displays the self-interaction potential strengths as a function of radius for the P250 simulation. The six subplots represent time-slices around the peak neutrino emission and each subplot displays the $\left(L_{\alpha},\left\langle E_{\alpha}\right\rangle\right)$ data valid for that time-slice as an inset table. The black line represents the matter potential which is $\sqrt{2}\,G_\text{F}\,Y_e\,\rho/m_N$, the blue and red lines are the electron and muon-or-tau self-interaction potentials respectively. The dashed blue and red lines represent the corresponding self-interaction potentials calculated via the BULB model. The gray horizontal lines represent the vacuum potentials with $H_\text{V,ij}=\Delta m_{ij}^2/(2\left\langle E\right\rangle)$ and the average energy being the mean across all flavors. From Fig. \ref{fig:Potentials} we conclude the self-interaction potentials are many orders of magnitude below the matter potential or vacuum Hamiltonian and indicates we need not consider neutrino self-interaction effects when calculating the the neutrino flavor evolution. As expected, the self-interaction potentials calculated via our shell model are lower than those calculated by the BULB model. Additionally, Fig. \ref{fig:Potentials} demonstrates how the shock affects the high density MSW resonance before affecting the low density MSW resonance. A similar analysis was performed for the P150 simulation and the neutrino self-interactions potential was even weaker than in the P250 simulation.


\subsection{Matter basis transformation probabilities\label{sec:PMatter}}

The matter basis is defined to be the basis where the eigenvalues of the total Hamiltonian $H= H_V +H_M$ appear on the diagonal. The eigenvalues are arranged so that they reflect the ordering of the masses appearing in the vacuum Hamiltonian. The matter basis for the antineutrinos is defined similarly. The significant advantage to using the matter basis over others is that for adiabatic evolution the evolution matrix in this basis is close to diagonal and transition probabilities $P_{ij}$ are constant. 

Indeed we find adiabatic evolution is almost exactly the case at early times because the density profiles of the simulations are not steeply falling functions of the radius $r$. This means the matter basis neutrino transition probabilities between states $\nu_i$ and $\nu_j$ are approximately unity for $i=j$ and zero for $i\neq j$. The same is also true for the antineutrinos. However, adiabatic evolution cannot continue for all epochs because the adiabaticity depends upon the density gradient and the density profiles from our simulation have moving shocks. The presence of shocks changes the adiabaticity and the change is most noticeable at those epochs where the shock is in the vicinity of the high (H) and/or low (L) Mikheyev-Smirnov-Wolfenstein (MSW) resonances \cite{2000PhRvD..62c3007D}. At these times there is near-total `hopping' from one matter eigenstate to the other \cite{1986PhRvL..57.1275P} which translates into the neutrino remaining in whichever flavor state it was in before the shock. Thus we expect significant changes in the flavor transition probabilities oscillations as the shock moves through the mantle of the supernova which will, in turn, lead to a change in the number of events we detect here on Earth. For a normal mass ordering both the H and L resonances affect the neutrinos because the L resonance is in the 12 mixing channel (by ij mixing channel we mean mixing is between matter states $\nu_{i}$ and $\nu_{j}$), and the H resonance is in the 23 channel. For an inverted mass ordering the L resonance remains in the 12 neutrino mixing channel but the H resonance moves to the 13 antineutrino mixing channel  \cite{2000PhRvD..62c3007D}. 

\begin{figure}[ht]
\includegraphics[trim={0 2.3cm 0 0},clip,width=0.75\linewidth]{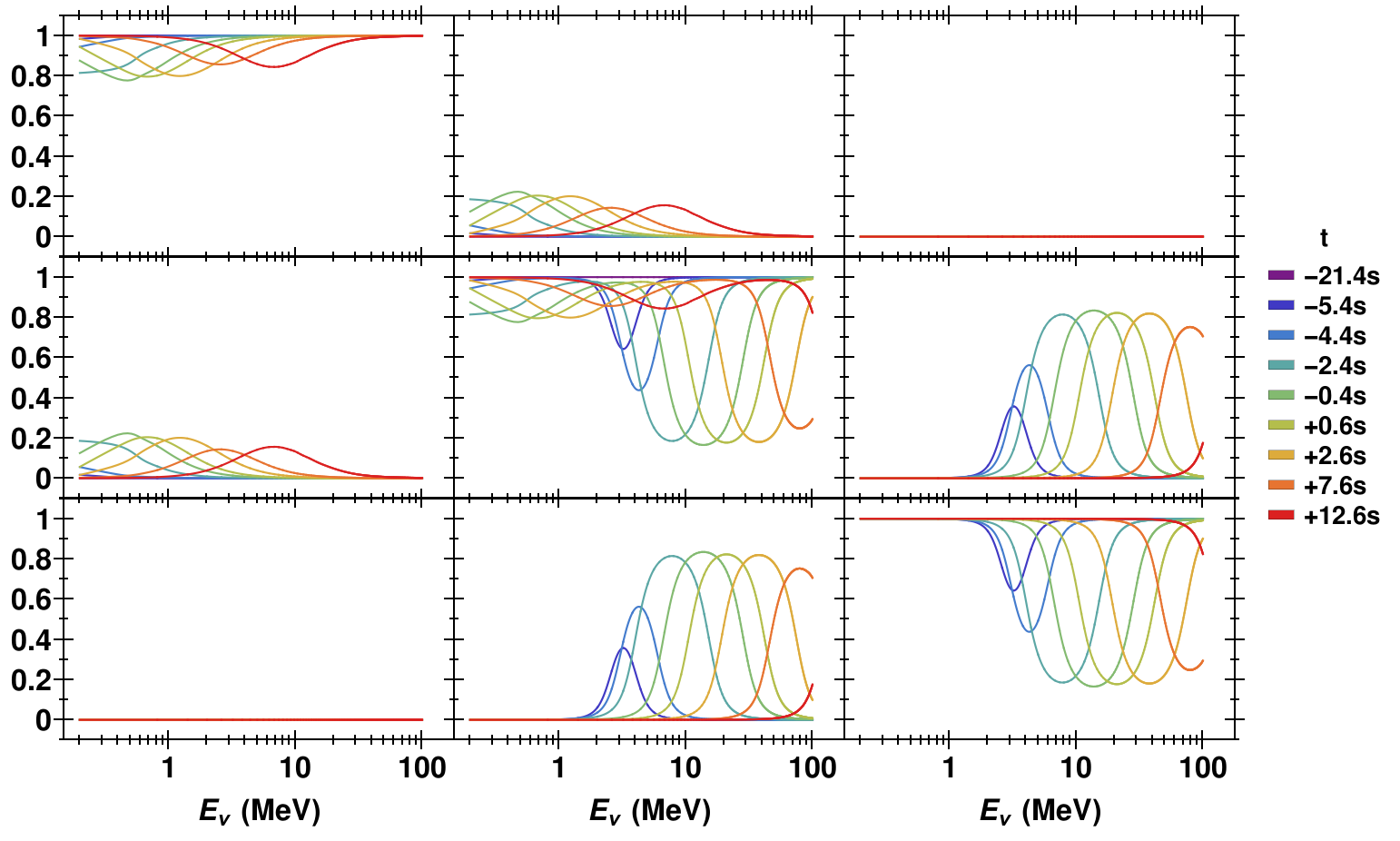}
\includegraphics[width=0.75\linewidth]{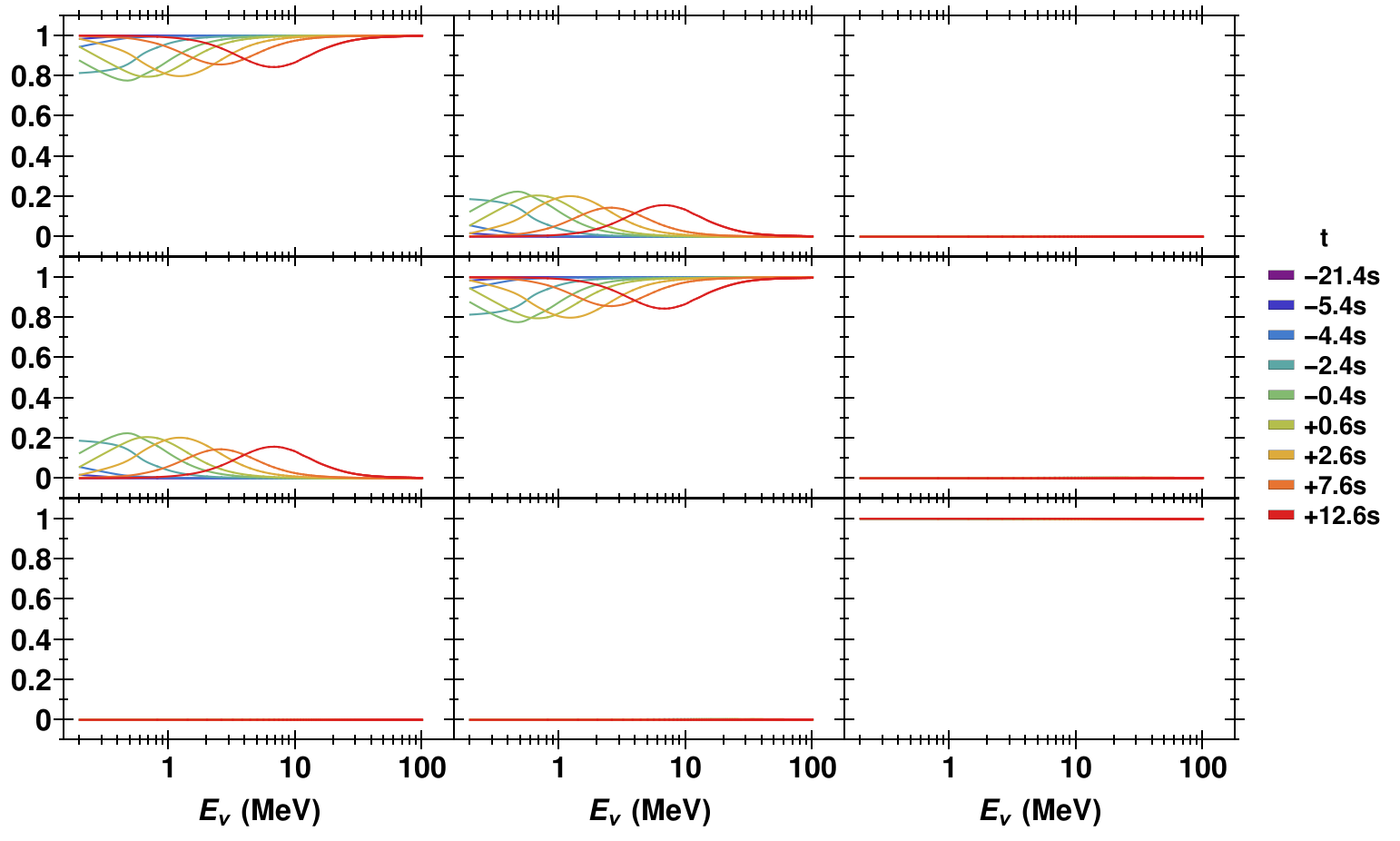}
\caption{The neutrino matter basis transition probabilities $P^\text{(mm)}_{ij}\left(E_\nu\right)$ as a function of energy. In both the subplots the top row, from left to right, shows $P^\text{(mm)}_{11}$, $P^\text{(mm)}_{12}$ and $P^\text{(mm)}_{13}$. The middle row from left to right shows $P^\text{(mm)}_{21}$, $P^\text{(mm)}_{22}$ and $P^\text{(mm)}_{23}$ and the bottom row from left to right shows $P^\text{(mm)}_{31}$, $P^\text{(mm)}_{32}$ and $P^\text{(mm)}_{33}$. The mass ordering is normal for the top subplot, inverted for the bottom subplot and the time is given in the legend. Results are for the P250 simulation.}
\label{fig:Sm}
\end{figure}

\begin{figure}[ht]
\includegraphics[trim={0 2.3cm 0 0},clip,width=0.75\linewidth]{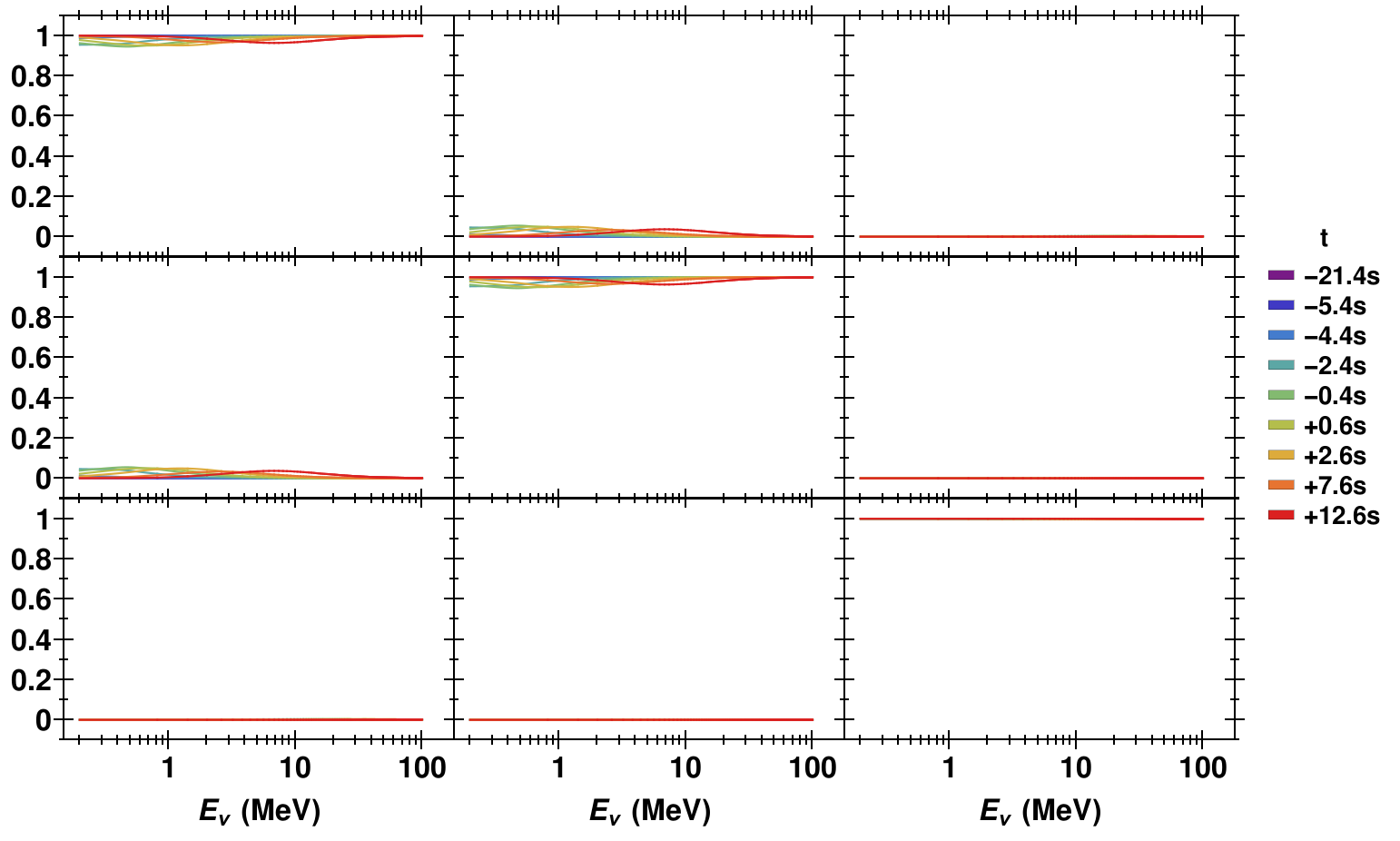}
\includegraphics[width=0.75\linewidth]{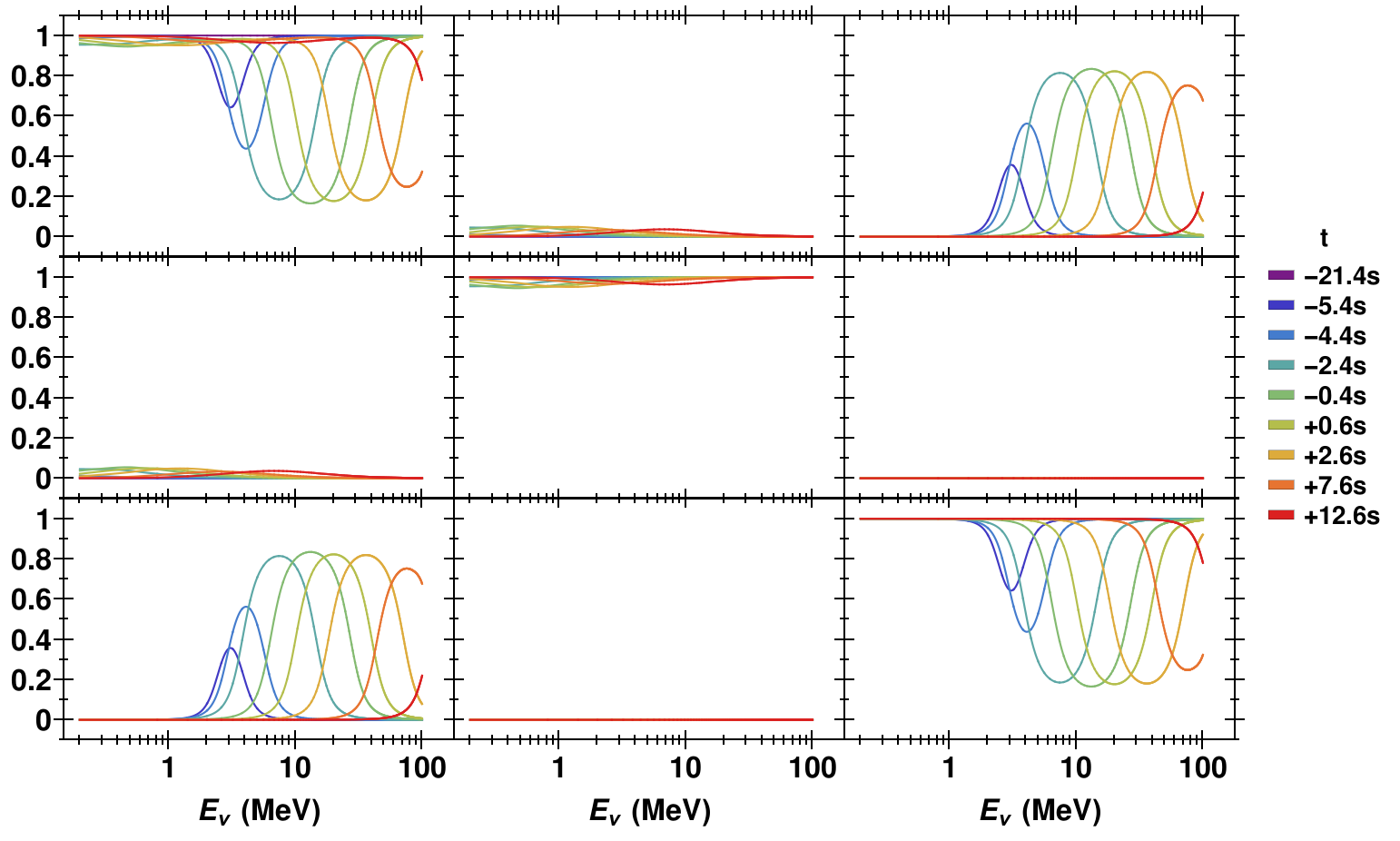}
\caption{The antineutrino matter basis transition probabilities $\bar{P}^\text{(mm)}_{ij}\left(E_\nu\right)$ as a function of energy. The layout is the same as Fig. \ref{fig:Sm}. Results are for the P250 simulation.}
\label{fig:SmBar}
\end{figure}

Figure \ref{fig:Sm} shows the neutrino matter basis transition probabilities $P^\text{(mm)}_{ij}\left(E_\nu\right)$ for both mass orderings and as a function of energy for a number of time-slices around peak neutrino emission for the P250 simulation. For the top panel, which represents normal mass ordering, we see energy-dependent departures from adiabatic behavior in the 12 and 23 mixing channels. This partial diabatic evolution appears at $t\sim-5.4$ s for neutrinos with energy around 0.2 MeV in the 12 mixing channel and around 3 MeV in the 23 mixing channel, and moves to higher energies at later times. The evolution of the transition probabilities are easily understood. As the SN evolves, the shock moves from high to low densities and the MSW resonances are inversely proportional to the neutrino energy. As the figures shows, the diabaticity in the 23 mixing channel ($\sim80\%$) is much higher that in the 12 mixing channel ($\sim20\%$) due to the smaller value of $\theta_{13}$ compared to $\theta_{12}$.

The bottom subplot of Fig. \ref{fig:Sm} represents neutrino matter basis transition probabilities for inverted mass ordering. Here we only expect diabatic behavior in the 12 mixing channel because only the L resonance occurs for neutrinos in inverted mass ordering. Indeed that is what is is visible in the bottom subplot and the energy and time dependence of this Low MSW resonance is the same as for the the top subplot.

Similarly, Fig. \ref{fig:SmBar} represents the antineutrino matter basis transition probabilities $\bar{P}^\text{(mm)}_{ij}\left(E_{\bar{\nu}}\right)$ as a function of energy for a number of time-slices around peak antineutrino emission for both mass orderings for the P250 simulation. Figure \ref{fig:SmBar} shows similar behavior as seen in Fig. \ref{fig:Sm}. For antineutrinos in the inverted mass ordering, the diabatic behavior is seen in the 13 mixing channel as was seen in the 23 mixing channel for neutrinos in the normal mass ordering. This happens because the H resonance switches to the antineutrinos for the inverted mass ordering and is between antineutrino matter states $\bar{\nu}_1$ and $\bar{\nu}_3$. A small amount of mixing is seen in the 12 antineutrino mixing channel in both the normal and inverted mass ordering but with much smaller amplitude. For the energies shown, the scale of the diabatic behavior for antineutrinos (in either mass ordering) in the 12 mixing channel is small ($<10\%$).


\subsection{Flavor basis transition probabilities at Earth\label{sec:PFlavor}}

\begin{figure}[ht]
\includegraphics[trim={0 3.2cm 0 0},clip,width=0.75\linewidth]{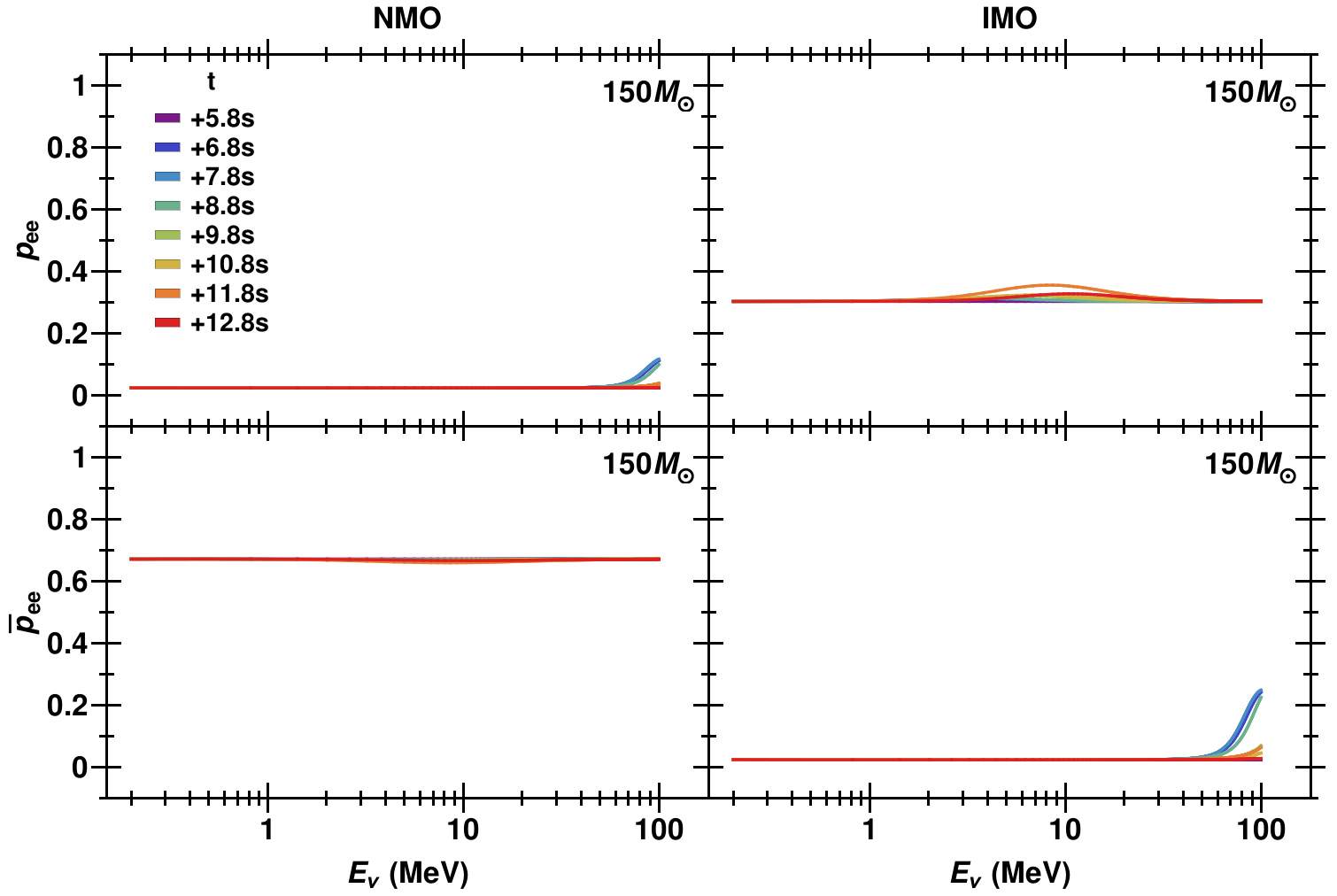}
\includegraphics[trim={0 0 0 1.2cm},clip,width=0.75\linewidth]{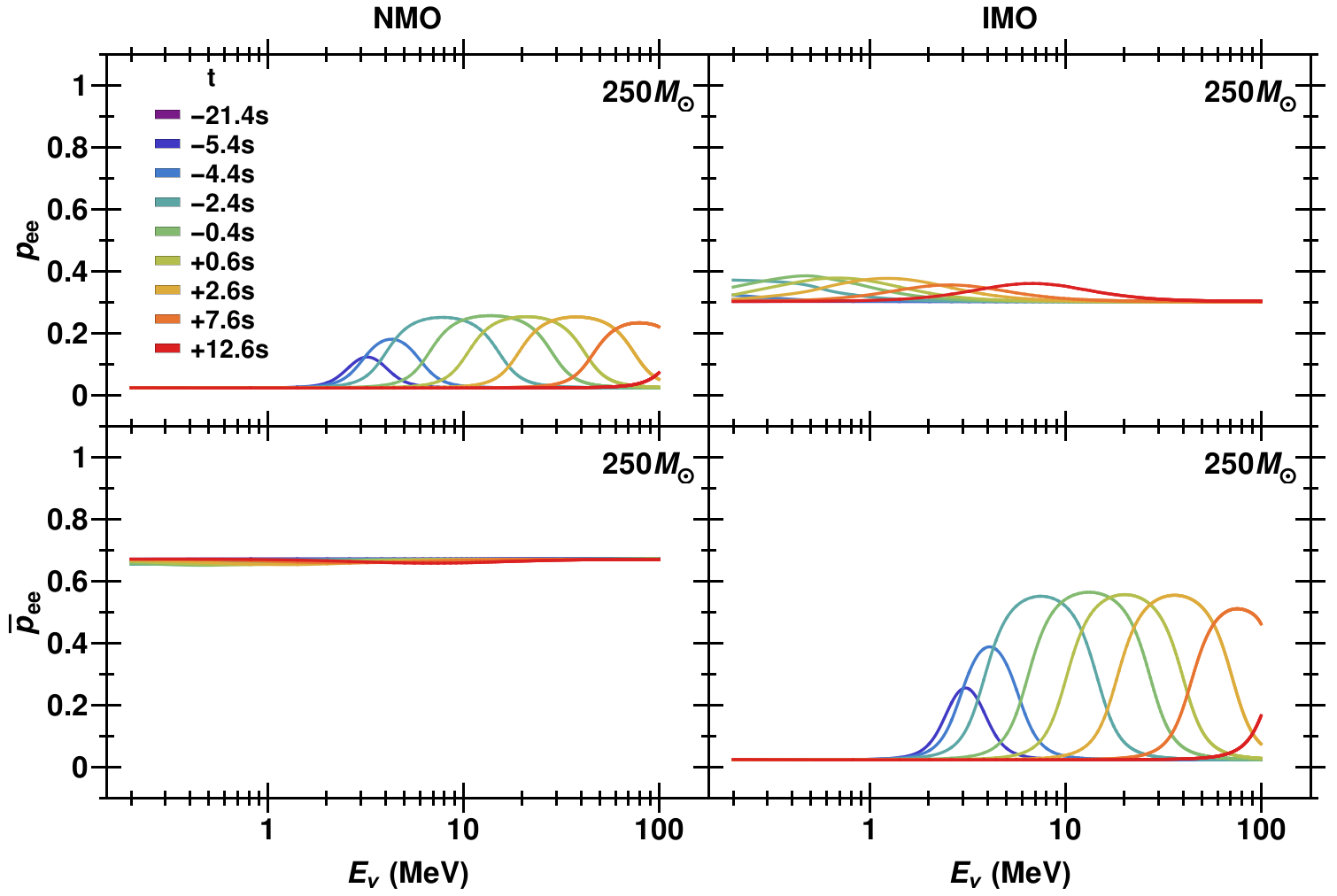}
\caption{$p_{ee}$ for neutrinos and $\bar{p}_{ee}$ antineutrinos and both mass orderings. The time slices chosen are representative of when oscillation features occur. The top four subplots are for the P150 simulation and the bottom four subplots are for the P250 simulation.
}
\label{fig:PeeAtEarth}
\end{figure}

Once the neutrinos have propagated through the star and start their long voyage to Earth, we need to account for the decoherence of the neutrino wavepacket \cite{1998PhLB..421..237G}. Accounting for this effect, the probability $p_{\alpha\beta}$ that a neutrino emitted as flavor $\beta$ in the supernova will be detected as flavor $\alpha$ at Earth is given by 
\begin{align}
p_{\alpha\beta}=\sum_i |U_{V,\alpha i} |^2 P^\text{(mf)}_{i\beta}\left(R_*,R_0\right),\label{eqn:Decoherence}
\end{align}
where $R_0$ represents the radius of the neutrino production point (near the center of the supernova which we take to be the origin) and $R_*$ represents the radius of the outer edge of the supernova. One defines antineutrino flavor basis survival probabilities $\bar{p}_{\alpha\beta}$ by using the matrix $\bar{P}^\text{(mf)}_{i\beta}\left(R_*,R_0\right)$. Note these probabilities $p$ and $\bar{p}$ are different from the transition probabilities we discussed in the previous section because they do not come from an evolution matrix. In Fig. \ref{fig:PeeAtEarth} the electron neutrino and antineutrino survival probabilities $p_{ee}$ and $\bar{p}_{ee}$ as a function of energy. Neutrino (antineutrino) survival probabilities are displayed in the top (bottom) row for each PISN simulation and the left (right) column represents the normal (inverted) mass ordering. It is clear the general trend is that flavor oscillations hamper $\nu_e$ or $\bar{\nu}_e$ survival. Only for $\bar{\nu}_e$ in NMO does greater that 50\% of the electron flavor survive. For the P250 simulation, the `humps' which appear for electron neutrinos in the NMO and electron antineutrino in the IMO are mostly caused by the passage of the shock through the H resonance. Without the shock the probability that an electron neutrino or electron antineutrino would be detected in the same state at Earth is just a few percent. As seen in the electron neutrinos in the IMO, the passage of the shock through the L resonance produces a much smaller effect and at later times. For the P150 simulation, it is clear that there are very few energy / time dependent oscillation effects. In this case the neutrino flavor evolution is almost entirely adiabatic at all times because the shock forms quite late and at low densities. In fact, for the P150 simulation, the shock never passed through the H resonance because it formed beyond that radius, and moved so slowly that it barely passes through the L resonance before the neutrino emission subsides.


\subsection{The neutrino flux at Earth\label{sec:OscFlux}}

\begin{figure}[ht]
\includegraphics[trim={0 2cm 0 0},clip,width=0.75\linewidth]{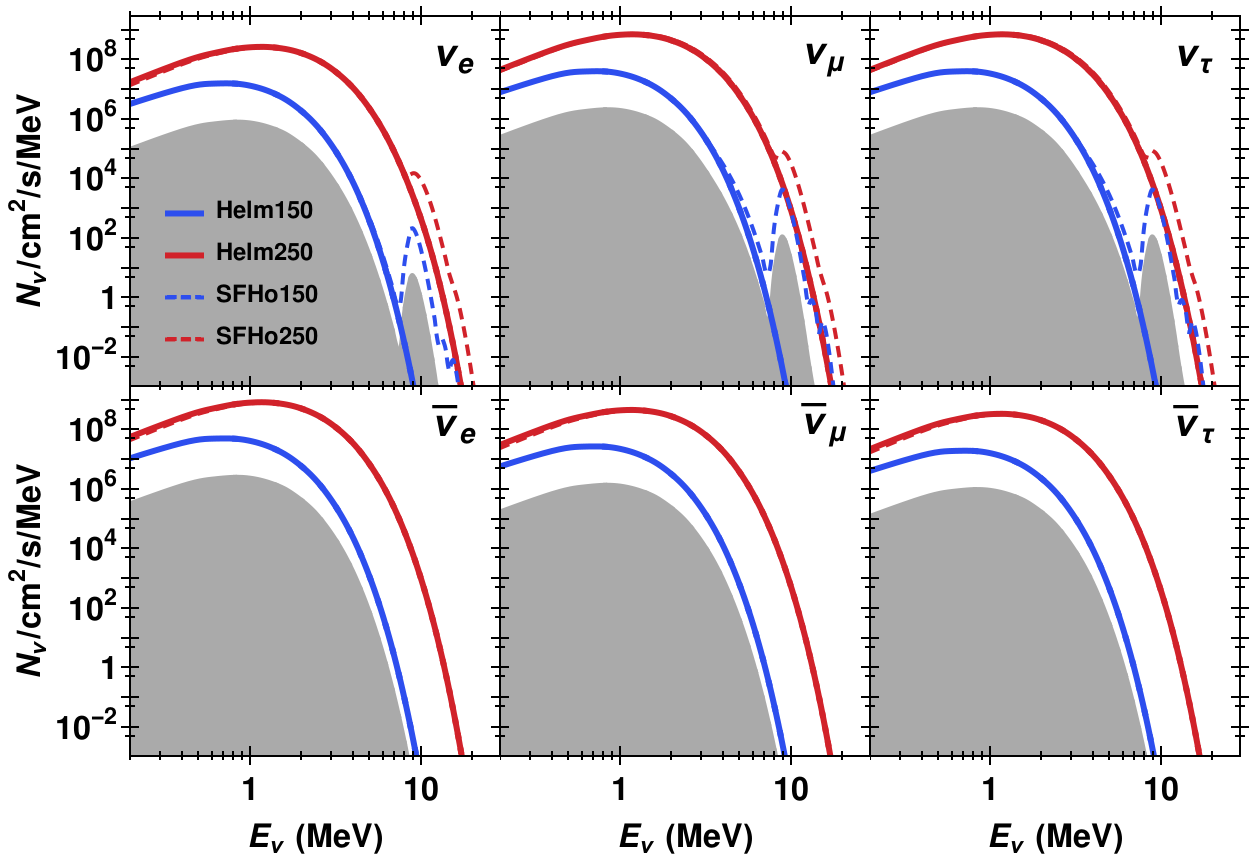}
\includegraphics[trim={0 0 0 0},clip,width=0.75\linewidth]{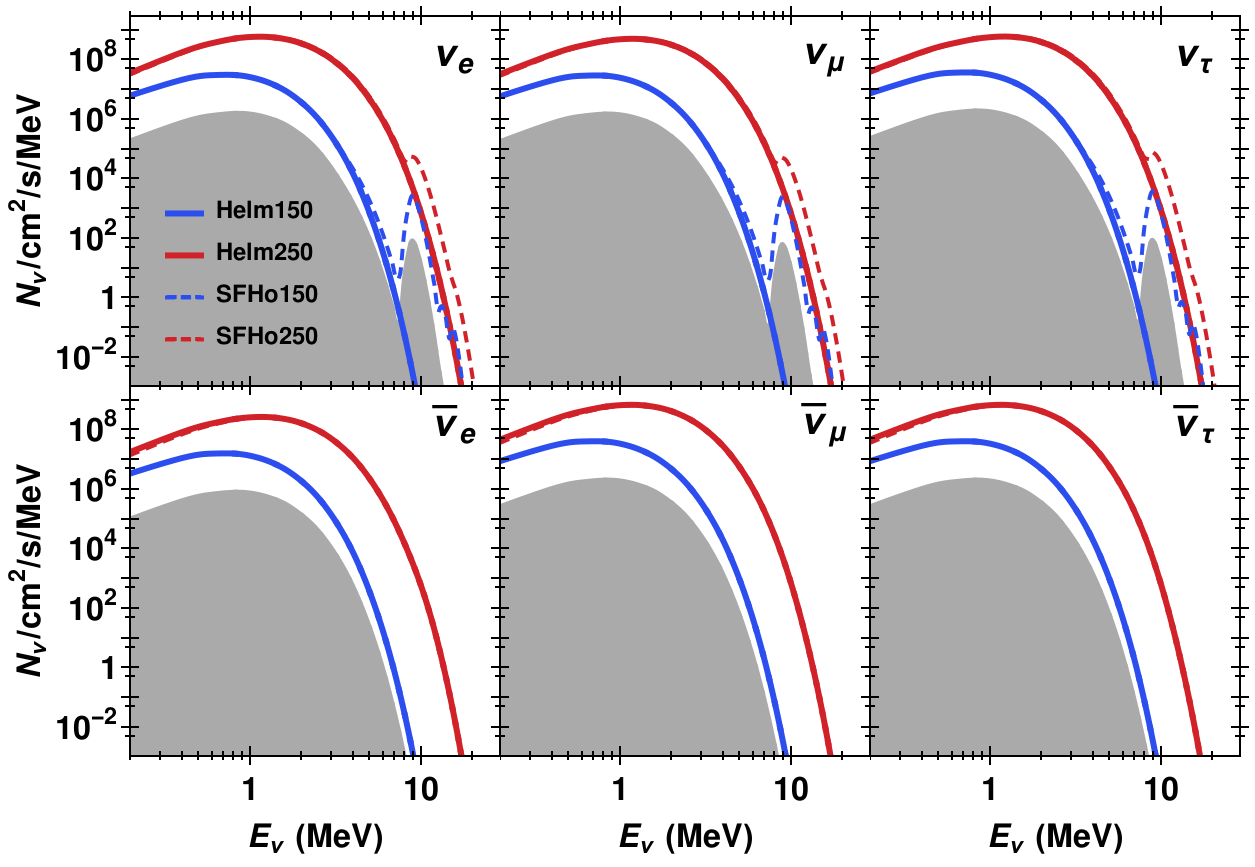}
\caption{Total oscillated neutrino flux from a PISN at 10 kpc. Each curve is the sum of all considered weak and thermal processes at the time slice of maximum emission. The top subplot is for NMO and the bottom subplot represents IMO. The gray region is the spectra from the SFHo, P250 results at $t=12.6$ s.}
\label{fig:OscFlux}
\end{figure}

By combining the flavor basis oscillation probabilities and the neutrino emission spectra, we calculate the neutrino flux seen by neutrino detectors at Earth. These fluxes are given by
\begin{align}
F_{\alpha} = \frac{1}{4\pi d^2}\sum_{\beta} p_{\alpha\beta}(E) \Phi_{\beta}(E)
\label{eqn:flux}
\end{align}
where $d$ is the distance to the supernova and $\Phi_{\beta}(E)$ is the differential spectrum of flavor $\beta$ at the point of emission. In what follows we set the distance to the PISN to be $d=10\;{\rm kpc}$ but will comment about the event rates at other distances. Figure \ref{fig:OscFlux} shows the flux as a function of neutrino energy including oscillations (at peak emission time). By comparing Fig. \ref{fig:OscFlux} and Fig. \ref{fig:UnOscFlux} the effects of neutrino oscillation can be easily seen. Once again, the red (blue) lines represent the P250 (P150) simulation and the solid and dashed lines represent the results using the Helmholtz and SFHo EOSs respectively. The gray region is the oscillated spectra from the SFHo, P250 results at $t=12.6$ s. In the SFHo results, the 10 MeV spectral feature mentioned in Sec. \ref{sec:Production}, which prior to oscillation was only present in the $\nu_e$ flux, is now present in all three neutrino flavors. Furthermore, in all cases, even though the unoscillated electron flavor flux dominates over the other flavors, the oscillated flux spectrum shows that much of this electron flavor has oscillated into muon and tau flavor in both the neutrino and the antineutrino case. The effects of the shock are difficult to see on the logarithmic scales used to make these figures, but they exist between $t=-5.4$ s and $t=0$ s for the P250 simulation. At these times the changes, due to diabatic evolution induced by the shock, occur at energies sufficiently close to the energy of peak emission to make an appreciable difference. All other diabatic effects of the shock occur at energies where the luminosity was low enough that any variation is unnoticeable.

\section{Neutrino Detection \label{sec:NeutrinoDetection}}
Ultimately, the goal is to be able to measure a neutrino signal in a neutrino detector and by using that signal, be able to identify source characteristics. In order to determine what the neutrino signal from our PISN simulations would be, we use the software package SNOwGLoBES to simulate the expected event rates in a variety of detectors. Table \ref{table:Detectors} lists the detectors considered together with their type and mass. The detector description includes the work "type" to indicate that the detector model used in SNOwGLoBES is only approximate. This is because, for the existing detectors, more accurate descriptions of the detectors are not openly published and for the future detectors, the detector specifications are not finalized. Furthermore, realistic detector thresholds are not included here because they are not well established for the future detectors and thus we do not include them for the existing detectors for consistency. All results include a computational 0.5 MeV threshold. We also assume perfect detection efficiency, that is, the incoming neutrino's energy is perfectly reconstructed from the detected particle's energy. This assumption is again made because efficiencies are not known for all the detectors considered and thus the events calculated here are labeled interaction events. These assumptions are overly optimistic, but they allow more consistent comparison between detectors given the information available. At the end of this section, an analysis of the effect of detector thresholds and efficiency (also called smearing) is presented.

\renewcommand{\arraystretch}{0.75}
\begin{table}[ht!!!]\centering
		\begin{tabular}{| l | c | c |}
			\cline{1-3}
			{\bf Detector}&{\bf Type}&{\bf Mass (kt)}\\ 
			\cline{1-3}
			Super-Kamiokande type: 30\% phototube coverage\cite{2003NIMPA.501..418F}$^*$& Water Cherenkov&50\\  
			Hyper-Kamiokande type\cite{Hyper-Kamiokande:2016dsw}& Water Cherenkov& 374\\ 
			DUNE type detector\cite{2016arXiv160102984A}&Liquid Ar&40\\ 
			JUNO type detector\cite{2016JPhG...43c0401A}&Scintillator&20 \\
   			IceCube\cite{2011IceCube}
            & Water Cherenkov&3500$^\dagger$\\   
			\cline{1-3}
		\end{tabular}
   \caption{A summary of the detector types. $^*$ See SNOwGLoBES documentation for discussion on phototube coverage. $^\dagger$ For IceCube, the mass given is the `effective' mass.}
	\label{table:Detectors}
\end{table}
\renewcommand{\arraystretch}{1}

Table \ref{table:Events} shows the expected interaction event count for all the detectors listed in Table \ref{table:Detectors}, except IceCube, for a PISN occurring at 10 kpc. The results are reported for both the P150 and P250 simulations, both EOSs and both mass orderings as well as the case of no neutrino oscillations for comparison. These interaction event counts are the total from the whole $\sim 30$ s neutrino burst as well as across all energies. For reference, the expected event rates in these same detectors for a Type Ia supernova at 10 kpc are roughly two orders of magnitude smaller \cite{Odrzywolek2011a,2016PhRvD..94b5026W,2017PhRvD..95d3006W} than the P250 case, while the event rates for a CCSN are approximately three orders of magnitude larger \cite{Mirizzi2015a}.
Firstly, we note that the effect of neutrino oscillations is to lower the event count in all detectors and all cases. This is because charged current interactions usually produce the strongest signals and the overall trend of oscillations is to convert electron flavor into muon and tau flavor which are detected via neutral current. Next, we note that for Super-Kamiokande (SK), Hyper-Kamiokande (HK), and JUNO there are more events for NMO than IMO whereas for DUNE, there are more events for IMO than for NMO. This is because SK, HK and JUNO are sensitive to the inverse beta decay (IBD) channel and DUNE is not. As shown in Figure \ref{fig:OscFlux}, the $\bar{\nu}_e$ flux is about an order of magnitude greater in NMO than in IMO. Our investigations (detailed below) reveal that it is the IBD channel that is responsible for why the NMO event count is greater than the IMO event count in the detectors that are sensitive to it. Next, Table \ref{table:Events} shows that SK, DUNE and JUNO would detect several and HK would detect several tens of interaction events over the span of the neutrino signal for a PISN at the upper end of the progenitor mass range. Such a detection would indicate high-temperature nuclear burning for an extended period of time. This allows us to easily distinguish a PISN from other types of supernovae: the neutrino signal of a CCSN reaches a peak luminosity during the neutronization burst which occurs $\sim 100\;{\rm ms}$ after core bounce \cite{2010A&A...517A..80F}, the entire duration of the neutrino emission from a Type Ia is $\sim 5\;{\rm s}$ at most \cite{Odrzywolek2011a,2016PhRvD..94b5026W,2017PhRvD..95d3006W}, while the PISN neutrino signal lats for $\sim 30$ seconds and takes $\sim 15$ seconds to reach peak luminosity. Even if other discriminators are not available, this temporal evolution alone would be a smoking-gun signature of a PISN and that some stars must end their lives in this way. The table also reveals that the differences between the predicted number of events by the two EOSs is minor. Lastly, it is clear that the P150 simulation would just barely be observable at $d=10\;{\rm kpc}$ and that the P150 simulation produces around 25 times fewer events than the P250 simulation.

\renewcommand{\arraystretch}{0.8}
\begin{table}\centering
		\begin{tabular}{| c | l | r | r | r | r | r | r |}
			\cline{1-8}
            \multirow{2}{*}{\bf Mass}&
            \multirow{2}{*}{\bf Detector}&
            \multicolumn{2}{c |}{\bf NMO}&
            \multicolumn{2}{c |}{\bf IMO}&
            \multicolumn{2}{c |}{\bf Unoscillated}\\
			\cline{3-8}
            &&Helm&SFHo&Helm&SFHo&Helm&\multicolumn{1}{c |}{SFHo}\\
			\cline{1-8}
			\multirow{4}{*}{P150}
            &Hyper-Kamiokande&$ 1.77 $&$ 1.78 $&$ 1.74$&$ 1.75 $&$ 3.02 $&$ 3.05$\\ 
			&Super-Kamiokande&$ 0.24 $&$ 0.24 $&$ 0.23$&$ 0.23 $&$ 0.40 $&$ 0.41$\\ 
			&DUNE&$             0.14 $&$ 0.14 $&$ 0.15$&$ 0.15 $&$ 0.25 $&$ 0.25$\\ 
			&JUNO&$             0.10 $&$ 0.10 $&$ 0.10$&$ 0.10 $&$ 0.17 $&$ 0.17$\\
			\cline{1-8}
			\multirow{4}{*}{P250}
            &Hyper-Kamiokande&$52.23 $&$50.08 $&$43.32$&$41.98 $&$85.70 $&$84.19$\\ 
			&Super-Kamiokande&$ 6.98 $&$ 6.69 $&$ 5.79$&$ 5.61 $&$11.46 $&$11.26$\\ 
			&DUNE&$             2.95 $&$ 2.78 $&$ 3.17$&$ 3.06 $&$ 5.30 $&$ 5.20$\\ 
			&JUNO&$             3.13 $&$ 3.00 $&$ 2.48$&$ 2.40 $&$ 5.06 $&$ 4.97$\\
			\cline{1-8}  
		\end{tabular}
   \caption{Numbers of interactions per detector for each mass ordering and a PISN at 10 kpc. These event counts are for the whole neutrino burst. The last two columns represents the number of interactions observed when neutrino oscillations are not taken into account.}
	\label{table:Events}
\end{table}
\renewcommand{\arraystretch}{1}

\begin{figure}[b]
\includegraphics[trim={0 2cm 0 0},clip,width=0.75\linewidth]{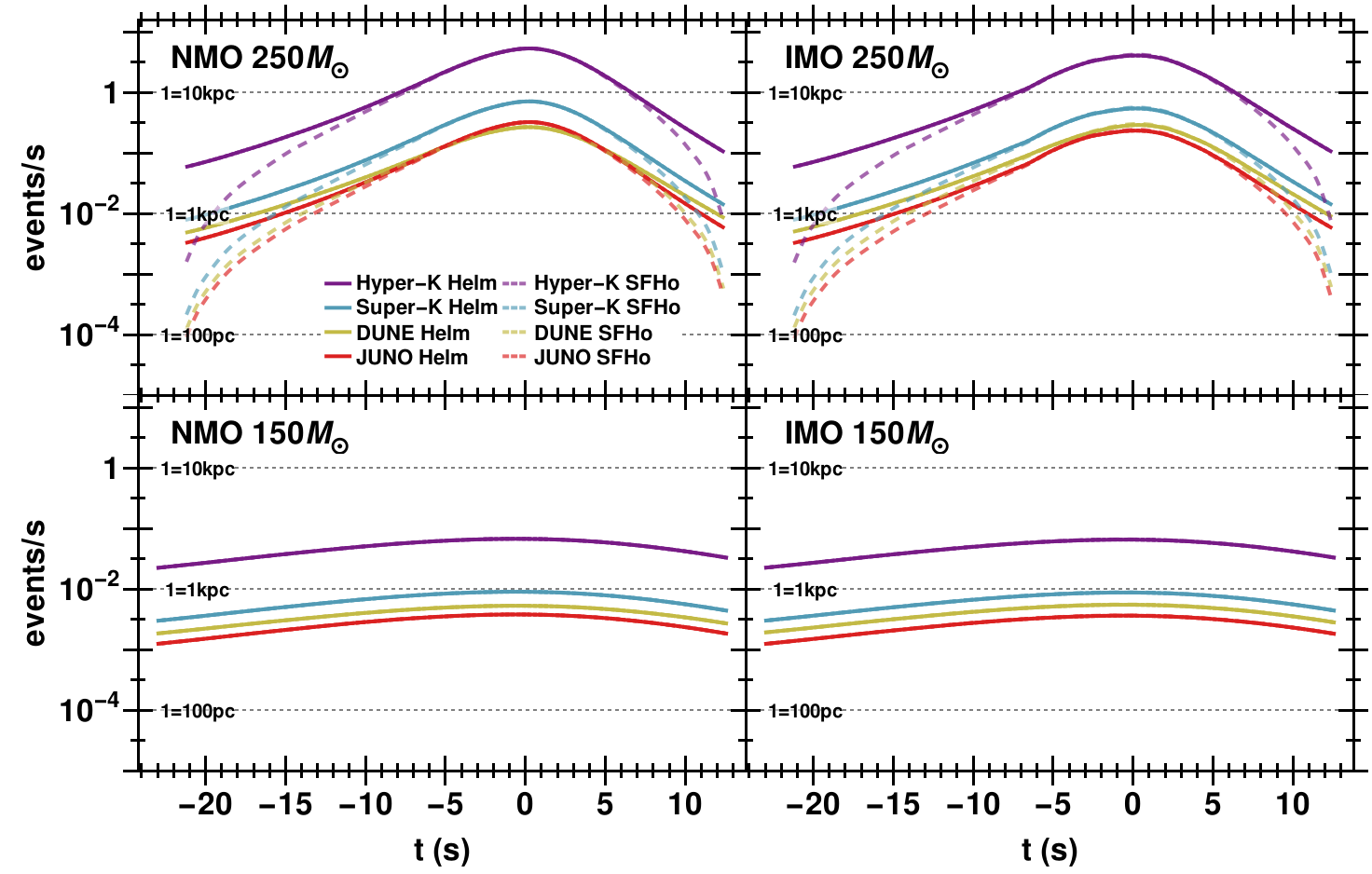}
\includegraphics[trim={0 0 0 0},clip,width=0.75\linewidth]{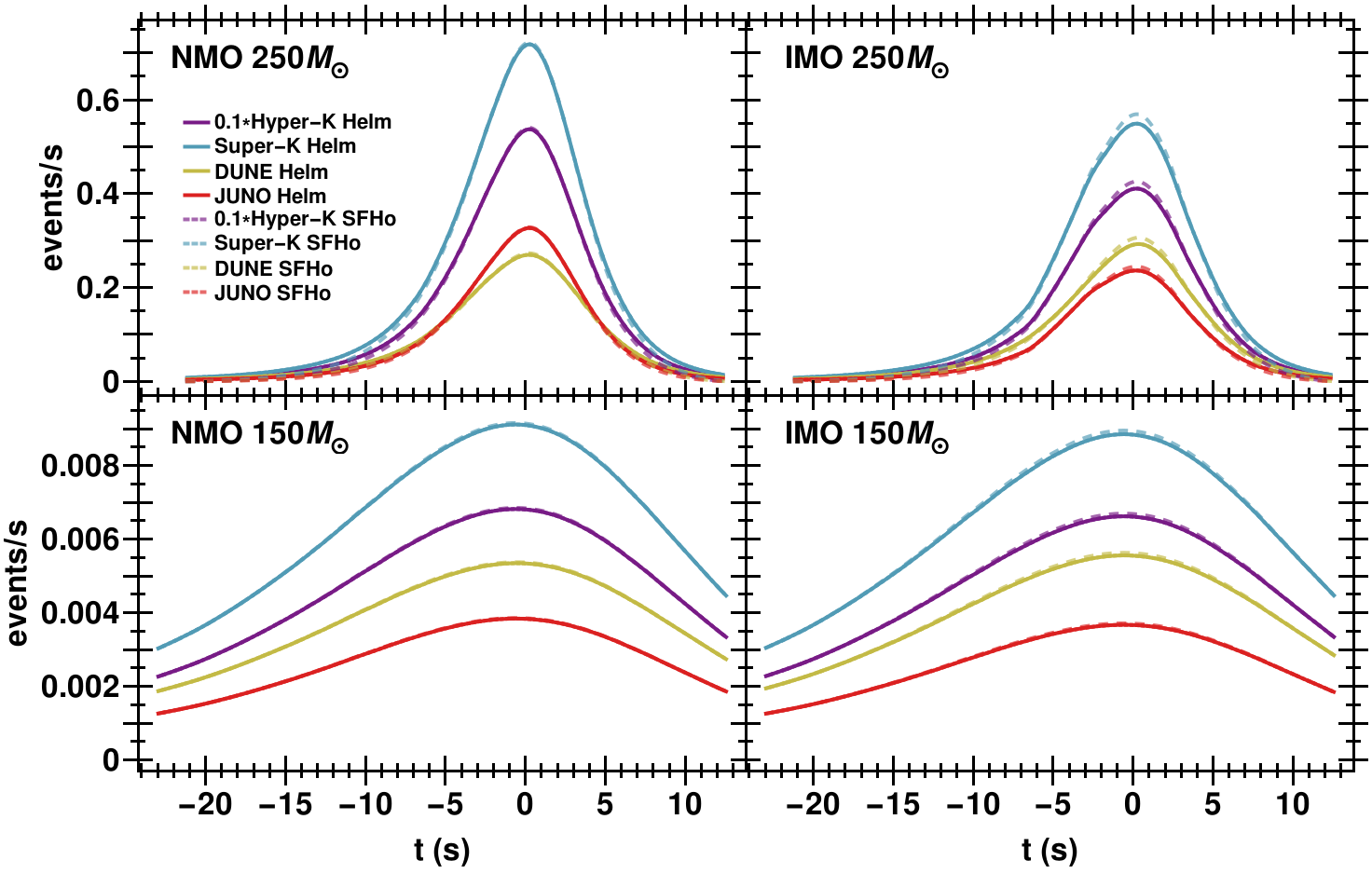}
\caption{Detector interaction event rate from a PISN at 10 kpc. The top four (bottom four) plots are on a log (linear) vertical axis. The left (right) plots are for NMO (IMO). For the log plots, the dashed horizontal lines show how the event rates would shift for a closer PISN. The purple line representing the event rate in HK in the bottom four plots has been rescaled to a tenth of its proper value for plotting convenience.}
\label{fig:EventsVsTime}
\end{figure}

By going beyond total event counts and considering the time and energy structure of the signal we can investigate the observability of the various temporal and spectral features in the neutrino signal. Figure \ref{fig:EventsVsTime} displays the interaction event rates expected to be observed in the detectors under consideration from a PISN at 10 kpc. The top four panels show that, in log scale, NMO and IMO give roughly the same event time-profile at all detectors however, the bottom four panels, using a linear scale, reveal the aforementioned differences between the rates from NMO and IMO. For both orderings our general expectation is for a Gaussian-like burst of neutrinos over a period of $\sim30$ s. Solid and dashed lines represent results from the Helmholtz and SFHo EOS respectively and it is clear that the choice of EOS has little impact. If we look very carefully in the P250 simulation we find in the IMO case a slight increase between $t=-5$ s and $t=0$ s due to shock induced diabatic evolution. This burst structure shows that at the peak, HK would be seeing a little more than 5 interaction events per second for the NMO. However, for SK, DUNE and JUNO, bigger time bins (or a nearer SN) would be required to confidently see activity per bin. As expected, the P150 simulation has a much wider and dimmer signal.

\begin{figure}[b]
\includegraphics[trim={0 2cm 0 0},clip,width=0.75\linewidth]{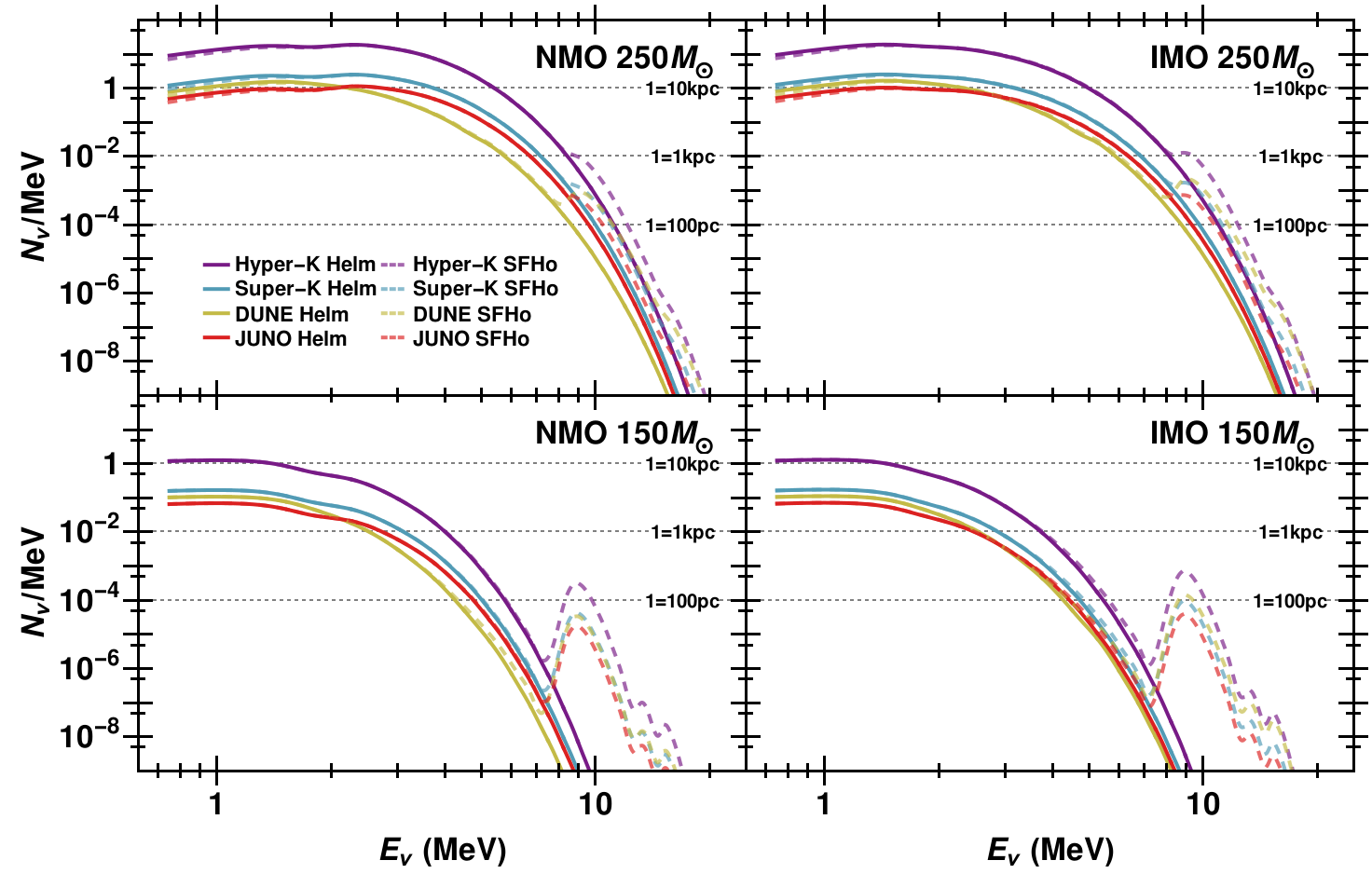}
\includegraphics[trim={0 0 0 0},clip,width=0.75\linewidth]{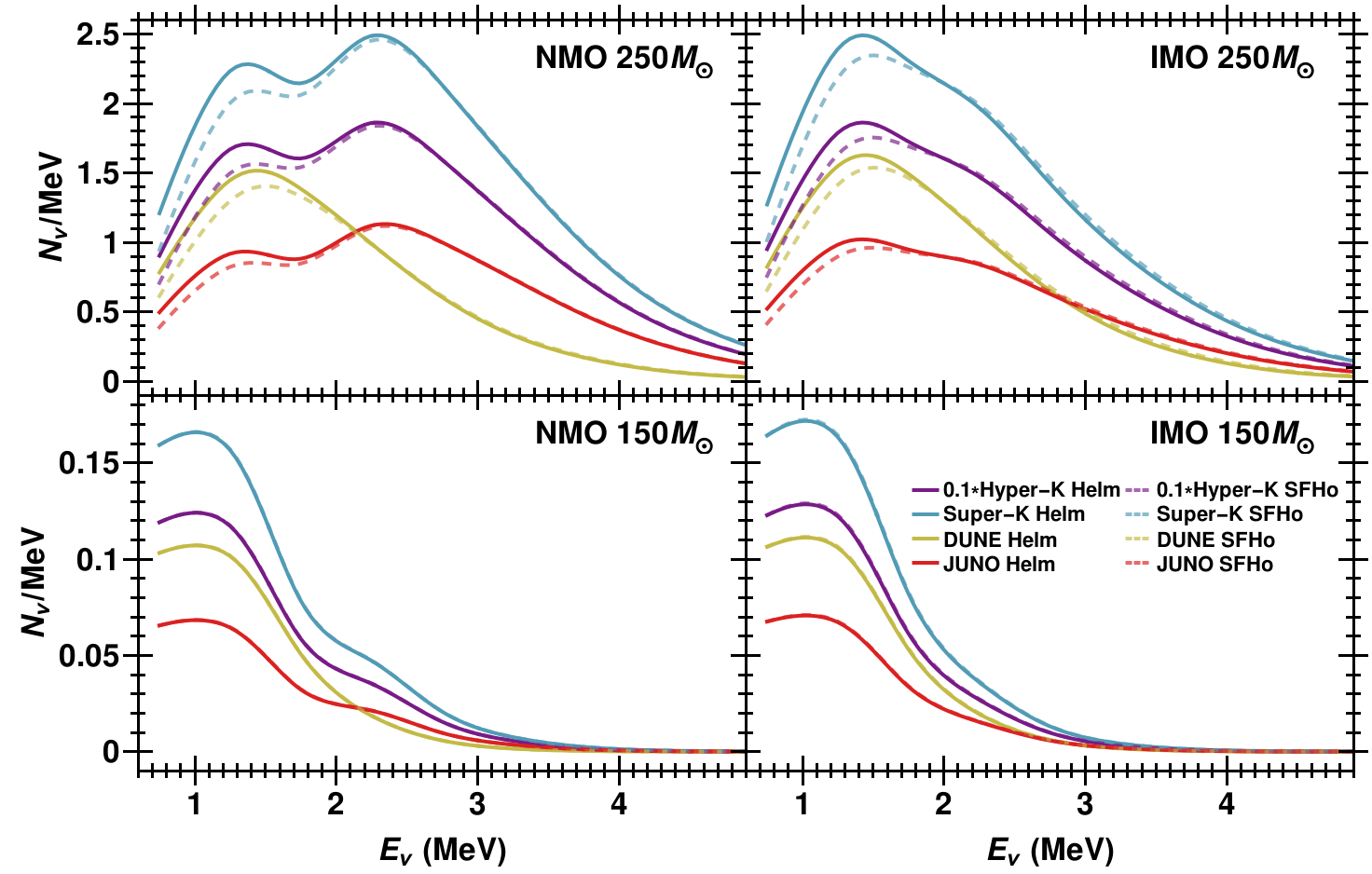}
\caption{Detector interaction event differential spectrum. Event count is the total for the whole neutrino burst. The top four (bottom four) plots are on log (linear) scales. The left (right) plots are for NMO (IMO). The horizontal lines in the top four plots show how the event rates would shift for a closer PISN. The purple line representing the event rate in HK in the bottom plot has been rescaled to a tenth of its proper value for plotting convenience.}
\label{fig:EventsDiffSpectrum}
\end{figure}

Figure \ref{fig:EventsDiffSpectrum} shows the energy structure of the neutrino burst released from a 10 kpc PISN as seen in the detectors under consideration. From all 8 plots, it is clear that the majority of the signal is below 5 MeV in all cases. The top plots reveals that even though the 10 MeV spectral feature from electron capture on copper is seen in both mass orderings using the SFHo EOS, it is a very small part of the signal and is thus less likely to be detected. The bottom 4 plots show a slight bump that is visible around 2-3 MeV in the NMO differential spectrum. This bump is from the contributions of the IBD channel which are more significant in NMO than in IMO. This is a very interesting spectral feature that, if ever detected at high enough statistical significance, could place constraints on the neutrino mass ordering. 

\begin{figure}[t]
\includegraphics[trim={0 0 0 0},clip,width=0.75\linewidth]{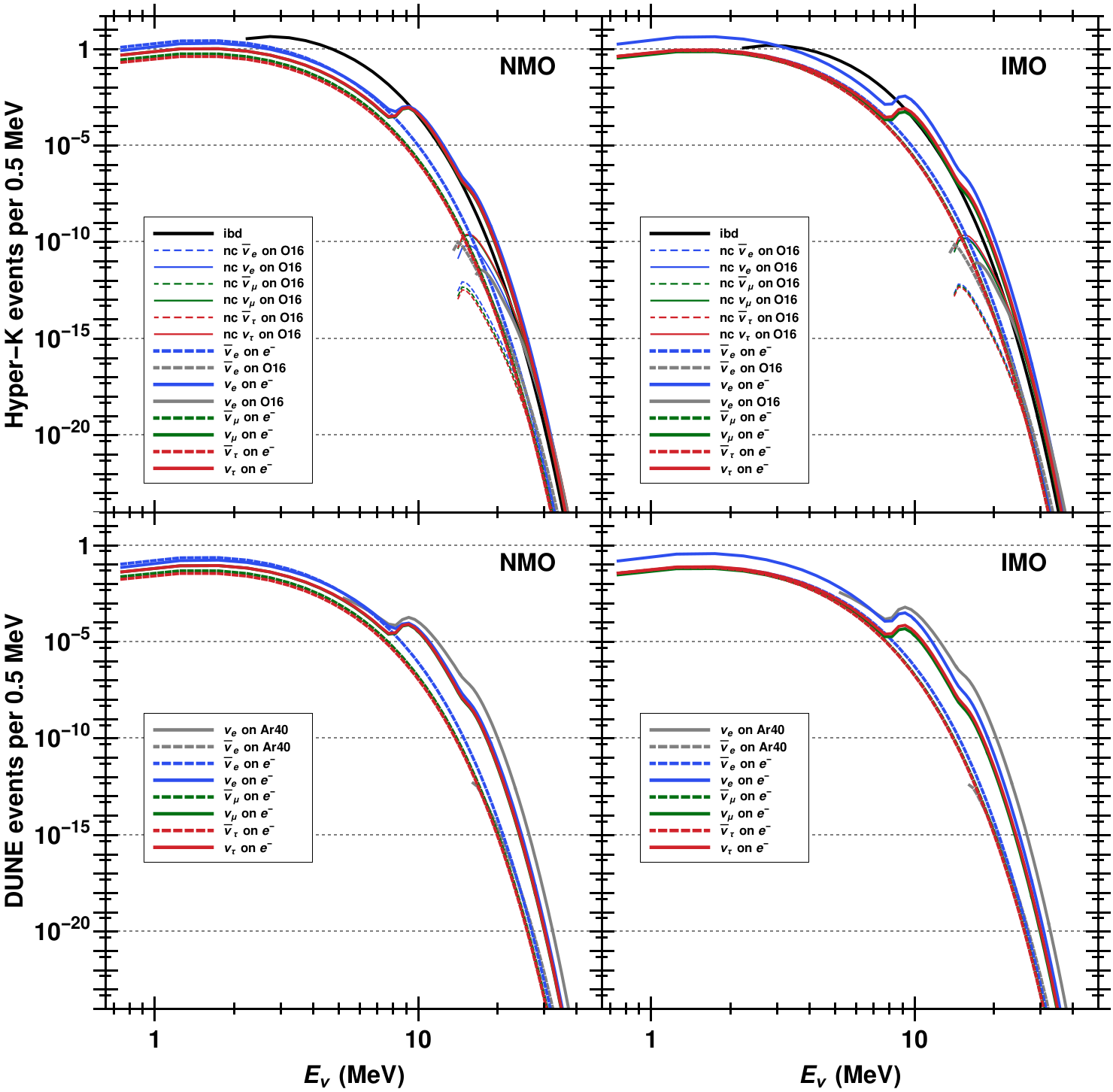}
\caption{Detector interaction event differential spectrum by channel. Event count is the total for the whole 30s neutrino burst. The top (bottom) plots are for HK (DUNE) and the left (right) plots are for NMO (IMO). Results are for the P250 simulation using the SFHo EOS.}
\label{fig:EventsByChannel}
\end{figure}

We now examine the spectral event rate per channel in order to quantify and illustrate the previous points about the strength of the IBD in NMO over IMO. Figure \ref{fig:EventsByChannel} shows the interaction event count per 0.5 MeV energy bins for the full $\sim30$ s duration of a P250 PISN at 10 kpc (using the SFHo EOS). The reason that the IBD channel is present for HK but not for DUNE is because it is dominant for detectors built from materials that are composed of hydrogen e.g. water and scintillator. As previously noted, Fig. \ref{fig:OscFlux} reveals that the $\bar{\nu}_e$ flux is about an order of magnitude greater in NMO than IMO. This is the reason why the IBD contribution is shown by Fig. \ref{fig:EventsByChannel} to be much larger in NMO than in IMO. This gives rise to the strong double peak seen in Fig. \ref{fig:EventsDiffSpectrum} for the water and scintillator detectors in NMO. Figure \ref{fig:EventsByChannel} also shows that the neutrino elastic scattering on free electrons form a significant contribution to the total event count. For HK, contributions from interactions with oxygen are quite minimal but for DUNE, contributions from interactions with argon are important, especially at energies above 4 MeV. Below $\sim8$ MeV, where the bulk of the signal is, Fig. \ref{fig:EventsByChannel} is unchanged by choice of EOS. Above $\sim8$ MeV, Helmholtz spectra continue to decline according to the thermal trends visible below $\sim8$ MeV.

\begin{figure}[ht]
\includegraphics[trim={0 0 0 0},clip,width=0.75\linewidth]{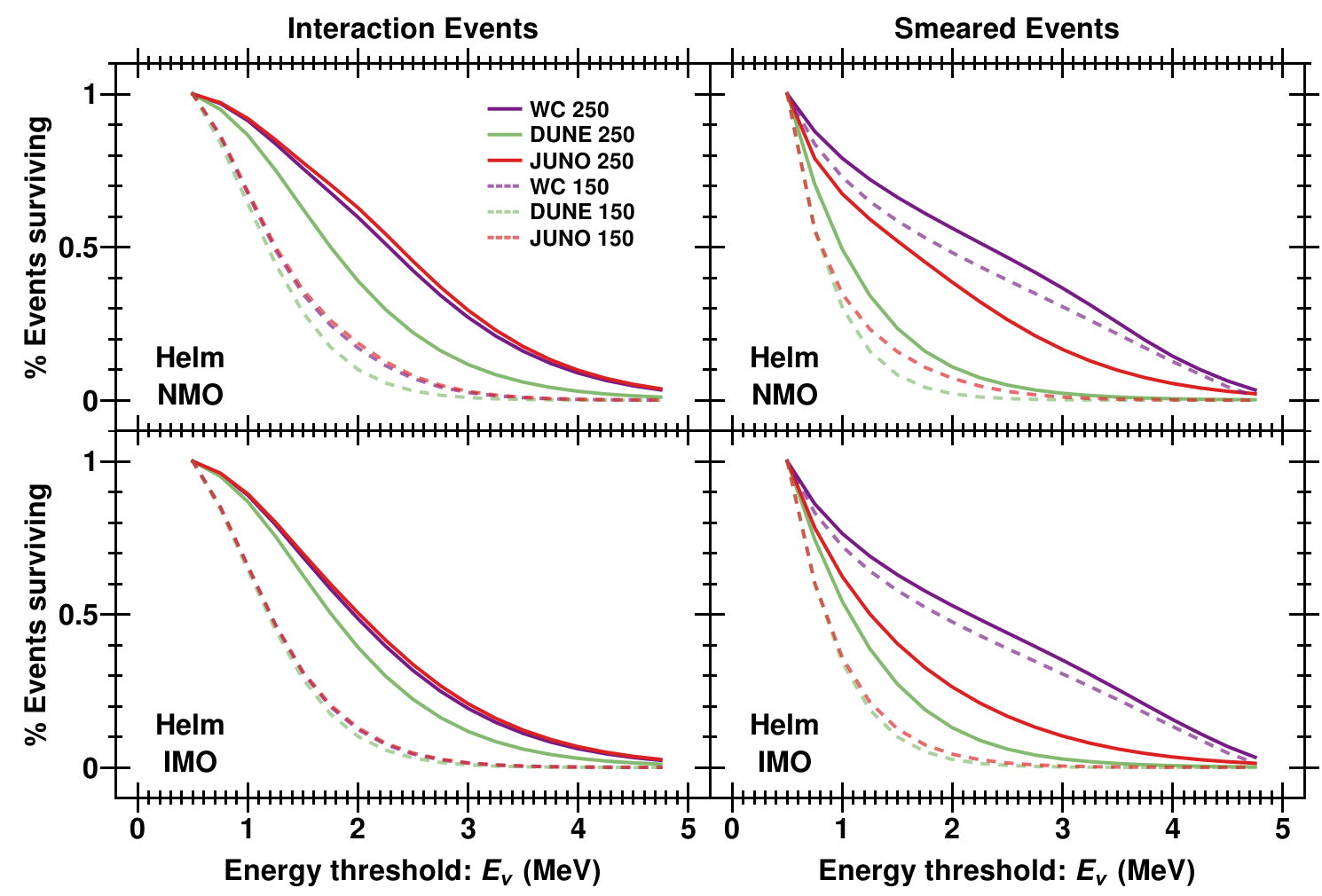}
\caption{Detector threshold analysis. The x-axis represents the simulated detector threshold level and the y-axis is the percentage of events that would be detected above this threshold. All results are for the full $\sim30$ s neutrino signal. The rows differentiate the two mass ordering cases. The left column is for interaction events and the right column is for smeared events according to the specifications employed by SNOwGLoBES. Results are presented for the Helmholtz EOS only. In our analysis, Super-K and Hyper-K have identical threshold structure and are here labeled as WC (for \underline{W}ater \underline{C}herenkov).}
\label{fig:ThresholdAnalysis}
\end{figure}

Finally, we examine the effects of detector thresholds and energy smearing which, considering that the bulk of the signal is below 5 MeV,  are expected to be an important consideration. Figure \ref{fig:ThresholdAnalysis} shows what percentage of events would survive given a particular detector event threshold. This analysis was for interaction events (left column), for energy-smeared events (right column), and for both mass orderings. The solid and dashed lines are for the P250 and P150 simulation respectively. The figure reveals that a threshold of 5 MeV would reduce the detectable signal in all cases essentially to zero. However, a threshold of 2 MeV means that, for the P250 simulation, Hyper-K, Super-K and JUNO would retain more than 50\% of their interaction events while DUNE would retain closer to 40\%. The case is worsened when energy-smearing is added but for a threshold of 2 MeV we see Hyper-K and Super-K still retain more than 50\% of their events, JUNO drops to 30-40\% (dependent on mass ordering) and DUNE drops to a bit more than 10\% retention. In short, energy smearing and detector thresholds do reduce the detectable signal for the P250 simulation, but given a threshold of around 2 MeV, much of the signal can still be detected, especially in the case of Hyper-K and Super-K. We also note that the spectral peak between 2-3 MeV which appeared for the NMO would still be visible for a 2 MeV threshold and, if observed, would still suggest the mass ordering is normal. The case of the P150 simulation however, shows that the Hyper-K and Super-K would still see 50\% of their smeared events, while for DUNE and JUNO, the signal is significantly attenuated.

\subsection{IceCube Event Rate}

The PISN neutrino signal in IceCube must be treated differently than the other detectors listed in Table \ref{table:Detectors}. The IceCube neutrino detector is located inside the ice that covers Antarctica and, while it is not primarily designed to investigate low energy neutrinos, it nonetheless is sensitive to them. These `detections' will consist of an increase of the background hum of low-energy events in the individual Photo-Multiplier Tubes. The events IceCube measures from a PISN are mostly from IBD but there is some contribution from elastic scattering off of electrons in the ice. However no direction or energy information of individual neutrino interactions will be extracted so the two event types cannot be distinguished. What can be determined is the overall flux of the neutrinos and the time structure. In order to be detected the PISN neutrino signal needs to be sufficiently above statistical fluctuations of the usual background event rate. Thus, even though the PISN will produce many events in IceCube, the metric of detectability is different from the other detector types considered. 
\begin{figure}[ht]
\includegraphics[trim={0 0 0 0},clip,width=0.45\linewidth]{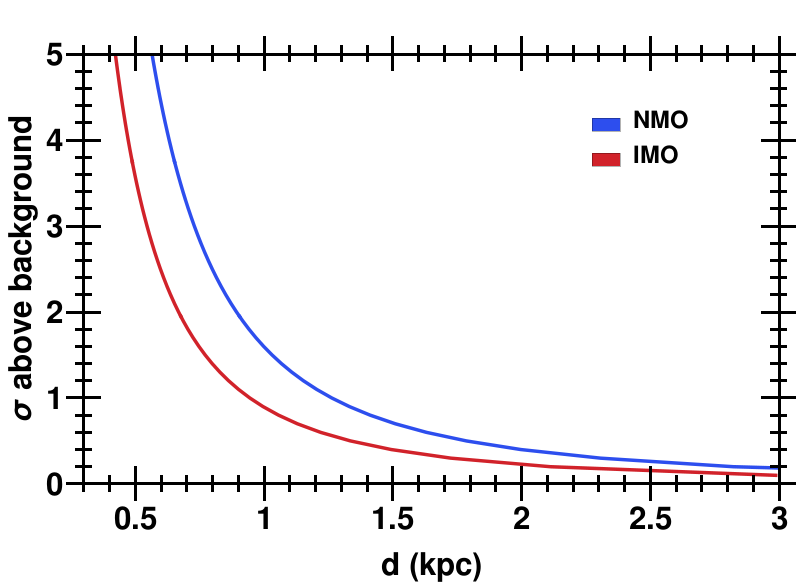}
\caption{The detection confidence of a PISN in the IceCube neutrino detector. The x-axis is the PISN distance and the y-axis is the confidence level of detection above background. Both neutrino mass orderings are represented. Results are for the P250 simulation. The choice of EOS has little impact in this analysis.}
\label{fig:IceCubeAnalysis}
\end{figure}
Figure \ref{fig:IceCubeAnalysis} shows the detection confidence of a P250 PISN at IceCube for a given PISN distance. As expected, given the abundance of $\bar{\nu}_e$ and hence IBD events, detection is easier in NMO than in IMO. However, in either mass ordering, it is clear that a PISN would have to be within one kiloparsec in order to be confidently detected above background at IceCube. The existence of stars with mass high enough to explode as a PISN at these relatively nearby distances seems quite unlikely. The P150 PISN case is even less likely to be detected by IceCube (3$\sigma$ requires the PISN to be within $\sim70$ pc).

\section{Astronomical context \label{sec:AstroContext}}

Thus far, we assumed that the PISN was located at 10 kpc. This was chosen because it is roughly the distance to the Galactic center and the distance used in the SN community as the most probable distance for the next Galactic SN. If we adopt a conservative lower limit for the ZAMS mass of a PISN of $100$~M$_{\odot}$ then this proximal distance is not unrealistically close: within the Milky Way there are several stars with masses around $100$~M$_{\odot}$ or above. These include:
\vspace{-\topsep}
\begin{enumerate}
\itemsep -0.3cm 
\item several stars in the Arches cluster \cite{1995AJ....109.1676N,1996ApJ...461..750C}, close to the Galactic center at a distance of $d\approx 8\;{\rm kpc}$ \cite{1993ARA&A..31..345R,2012arXiv1202.6128M,2017arXiv170801310C}, are estimated to have masses above $100$~M$_{\odot}$ \cite{2011A&A...535A..56G},
\item the OB association Cygnus OB2 \cite{1991AJ....101.1408M} at a distance of $d= 1.7\;{\rm kpc}$ \cite{0004-637X-699-2-1423} also contains stars with masses approaching $100$~M$_{\odot}$,
\item the mass of the primary in the HD 15558 system, at a distance of $d = 2.3\;{\rm kpc}$, was estimated to be $150 \pm 50$~M$_{\odot}$ \cite{2006A&A...456.1121D} (see also \cite{1538-3873-93-554-500}),
\item the primary star in the binary $\eta$ Carina which is, coincidentally, also at a distance of $d=2.3\;{\rm kpc}$, has a mass that is also modeled as being greater than $100$~M$_{\odot}$ \cite{2010ApJ...723..602K,2015MNRAS.447.2445C}.
\end{enumerate}
\vspace{-\topsep}
Whether these Galactic very high mass stars are too metal rich to explode as a PISN will have to be determined when the event occurs. The neutrino signal will easily distinguish the explosion type. A number of very high mass stars are also known within the Large Magellanic Cloud (LMC) - which has a lower metallicity of  $0.43\,$Z$_{\odot}$ \footnote{Computed from $\log( Z/Z_{\odot} ) = [\mathrm{Fe}/\mathrm{H}]$ and the values of [Fe/H] found in \cite{2016MNRAS.455.1855C} } - and which is at a distance of $d=49.97\;{\rm kpc}$ \cite{2013Natur.495...76P}: 
\vspace{-\topsep}
\begin{enumerate}
\itemsep -0.3cm 
\item The R136 open cluster contains nine stars with masses greater than $100$~M$_{\odot}$ at 1-sigma \cite{2016MNRAS.458..624C}.
\item NGC 3603-A1 with the more massive component of the binary having a mass $116 \pm 31$~M$_{\odot}$ \cite{2008MNRAS.389L..38S}.
\item WR21a whose more massive component has a mass of $103.6\pm 10.2$~M$_{\odot}$ \cite{2016MNRAS.455.1275T}.
\end{enumerate}
\vspace{-\topsep}
For stars at this distance the event rates given in table \ref{table:Events} need to be scaled downward by a factor of $25$. Thus we see Hyper-Kamiokande is the only detector at the present time or in the near future capable of detecting events from a large PISN at LMC distances. Again, the neutrinos from the explosion of any of these stars can be used to distinguish a PISN explosion from a core-collapse supernova. 

\section{Conclusion \label{sec:Conclusion}}

Pair-Instability Supernovae represent a intriguing option for the conclusion of stellar evolution in the case of very massive stars. They could also potentially be the source of some observed super-luminous supernovae and early generations of such stars would greatly change the chemical evolution and dynamics of galaxies. Much about PISN and their progenitors remains uncertain and observations of PISN would greatly improve our understanding. Distinguishing a PISN from other supernova types may not be as straight-forward as expected if the lightcurves of PISN at higher metallicities are not super-luminous but rather similar to the light curves of other supernovae. The goal of this paper is to determine whether the neutrino signal from a PISN in the Milky Way or a nearby galaxy could be detected and used to identify the explosion mechanism independent of the electromagnetic emission. 

The neutrino emission from a PISN simulation using P150 and P250 progenitors was calculated using the \textsc{NuLib} code. We determined that the dominant emission process was the thermal process of electron positron annihilation into neutrino pairs of all flavors. Also of great importance was the emission via weak processes which produced a 10 MeV peak in the spectrum that is familiar from SNe Ia investigations as coming predominantly from electron capture on copper. The neutrino emission is significant for a duration of $\sim30$ s, peaks at the time of maximum stellar compression, and the average energy is around 1-2 MeV.

We calculated the neutrino flavor oscillations through the stellar envelope and accounted for decoherence between the SN and Earth. The overall effect of flavor oscillations is to convert most of the electron neutrinos into muon and tau neutrinos. Electron antineutrinos have a greater probability of surviving, especially in NMO. The presence of the shocks in the SN density profile caused time/energy dependent diabatic evolution, which at certain times and energies, had a non-trivial impact on the oscillated flux.

The oscillated flux was then used as input for the code SNOwGLoBES in order to calculate the interaction event rate at various detectors. For a P250 PISN at 10 kpc, we find that HK could measure several tens of events and SK, DUNE and JUNO would measure several events. For the P150 PISN at 10 kpc only HK would detect enough events to observe the explosion. These predictions are not sensitive to the choice of EOS. Thus present and near-future neutrinos detectors can identify a PISN at the Galactic center providing a useful discriminator of the explosion type. The spectral distribution of the events reveals that most events would be below 5 MeV. However, because of the IBD contribution, the spectral distribution of events for the NMO has a double peak structure while the spectral distribution of events for the IMO does not. 

Our conclusion is that the gross features of the neutrino signal from a PISN are well-understood, and that the signal contains distinct signatures which are potentially detectable with present neutrino detectors should such a supernova occur in the Milky Way, and with Hyper-Kamiokande if a high mass PISN is located in the Large Magellanic Cloud. Additionally, the signal has spectral features that could help determine the neutrino mass ordering if the supernova is sufficiently close. Further refinement of the model and consideration of the small neglected effects will reduce the uncertainty of the predictions and allow for better extraction of quantitative information should a nearby PISN occur.



\section{Acknowledgments}
We are grateful to Evan O'Connor for his help with the implementation of \textsc{NuLib}.
This work was supported at NC State by DOE grants DE-FG02-02ER41216 and SC0010263.


\bibliographystyle{apsrev4-1}
\bibliography{main}

\end{document}